\documentclass[11pt,preprint,graphicx]{aastex}
\usepackage{graphicx}

\def\etal{\it et al. \rm }

\begin{document} 

\title{The Structure of Galaxies: III. Two Structural Families of Ellipticals}

\author{James M. Schombert$^{A,B}$}
\affil{$^A$Department of Physics, University of Oregon, Eugene, OR USA 97403}
\affil{$^B$jschombe@uoregon.edu}

\begin{abstract}

\noindent Using isophotal radius correlations for a sample of 2MASS ellipticals,
we have constructed a series of template surface brightness profiles to describe the
profile shapes of ellipticals as a function of luminosity.  The templates are a
smooth function of luminosity, yet are not adequately matched to any fitting
function supporting the view that ellipticals are weakly non-homologous with respect
to structure.  Through comparison to the templates, it is discovered that ellipticals
are divided into two families; those well matched to the templates and a second class
of ellipticals with distinctly shallower profile slopes.  We refer to these second
type of ellipticals as D class, an old morphological designation acknowledging 
diffuse appearance on photographic material.  D ellipticals cover the same range of
luminosity, size and kinematics as normal ellipticals, but maintain a signature of
recent equal mass dry mergers.  We propose that normal ellipticals grow after an
initial dissipation formation era by accretion of low mass companions as outlined in
hierarchical formation scenarios, while D ellipticals are the result of later equal
mass mergers producing shallow luminosity profiles.

\end{abstract}

\section{Introduction}

Since the time of Hubble, elliptical galaxies have been our purest morphology form.
Having only minor irregularities to their Keplerian shaped isophotes, ellipticals
distinguish themselves in their ease of classification and high repeatability in
subjective morphological schemes.  The brightest galaxies in the Universe are 
ellipticals, often located in the densest environments, making them well studied
signposts to high redshift and critical test particles to scenarios of galaxy
formation and evolution.

Uniformity in morphology and color for ellipticals suggests a simpler history of
evolution than other galaxy types, especially in relating luminosity to stellar mass
without the complications of ongoing star formation.  This scenario is supported by
the well defined relationships between luminosity and kinematics (the Fundamental
Plane, FP, Djorgovski \& Davis 1987; Burstein \etal 1997), the most precise relationship
found for galaxies.  While the homogeneous nature of ellipticals has been used to
argue for uniform, and early, formation processes (Tantalo, Chiosi \& Bressan 1998),
observations of high redshift ellipticals presents a more complicated picture of
stochastic mergers (Kauffmann \& Charlot 1998) that should reflect into present-day
structure.  And, structural non-homology has been argued to be one of the primary
reasons for non-linearity or a tilt to the FP (Hjorth \& Madsen 1995; Graham \&
Colless 1997).

The regularity in structure with luminosity for present-day ellipticals was enhanced
by the discovery of a number of scaling laws such as the one between color-magnitude
(Visvanathan \& Sandage 1977), the Kormendy relation (Kormendy 1977) and
luminosity-velocity dispersion (Faber \& Jackson 1976).  In terms of structure, the
highly uniform Keplerian shape to isophotes of ellipticals (Jedrzejewski 1987) allows
for the parameterization of elliptical structure into a few simple variables.  The
success of resolving elliptical structure was demonstrated by the `Photometric
Plane', a version of the Fundamental Plane that uses only luminosity and structural
information to characterize ellipticals (Graham 2002).

The study of the structure of ellipticals has become particularly salient in the last
decade for theoretical frameworks (e.g., hierarchical CDM) which have been successful
in explaining large-scale structure and provided an accurate prediction to galaxy
structure (Driver 2010).  Dividing galaxies by their structure also isolates features
that reflects formation history versus components that have evolved with time.  In
addition, measurable structure features are a valuable commodity for modelers, and
their simulations, to connect observables to physical processes in order to discovery
a few physical parameters which explain the range of galaxy properties and
morphology.

The best method for studying structure in ellipticals is through the analysis of
their surface brightness profiles, the run of isophotal luminosity with radius.  In
addition to 1D luminosity profiles, various 2D structure parameters have also been
defined (such as concentration, asymmetry and clumpiness, see Conselice 2003) which
are extremely useful in developing quantitative morphology.  However, as early-type
galaxy isophotes are typically elliptical (ignoring small boxy and disky
perturbations), these 2D isophotes can easily be reduced to a 1D surface brightness
profile.  These 1D profiles can than be further reduced to a few parameters by
matching the profile to an algebraic fitting function.  Typical fitting functions
will have resulting parameters that represent a characteristic surface brightness,
characteristic scale size and profile mean slope for each galaxy.

Many fitting functions have been proposed, and used, in past studies (see Graham 2013
for a review).  The two most popular (i.e., producing the most homogeneous
relationships between galaxy types) are the $r^{1/4}$ law (de Vaucouleurs 1948) and
the S\'{e}rsic $r^{1/n}$ model (S\'{e}rsic 1963).  The $r^{1/4}$ law is by far the
simplest (two variables which, if correlated, leads to structural homology), but is
clearly inadequate for describing an elliptical from core to halo (Schombert 1986).
The S\'{e}rsic $r^{1/n}$ model is very useful for ellipticals with resolved cores
(Graham 2002) and provides an additional shape parameter (the $n$ index) beyond the
$r^{1/4}$ law, but also has deficiencies for elliptical halos (Schombert 2013).

To summarize the results from Schombert (2013), it was found that the S\'{e}rsic
$r^{1/n}$ model produced good fits to the core regions of ellipticals ($r <
r_{half}$), but fails for the entire profile of an elliptical, i.e. from core to
halo, due to the competing effects on the S\'{e}rsic $n$ index and the fact that the
interior shape of an elliptical is only weakly correlated with its halo shape. In
addition, it was found that there exists a wide range of S\'{e}rsic parameters that
will equally describe the shape of the outer profile (i.e., $n$ becomes degenerate at
large values), degrading the S\'{e}rsic model's usefulness as a describer of the
entire profile.

Empirically determined parameters, such as half-light radius and total luminosity,
were found to have much less scatter than fitting function variables, begging the
question on why fitting functions are applied in the first case.  To this end, this
paper presents a description of the structure of ellipticals using template profiles
empirically derived from the 2MASS $J$ images of over 300 ellipticals.  This follows
the prescription from Schombert (1986) in making $V$ templates for the study of D and
cD galaxies.  The unexpected consequence of the template construction was the
discovery that ellipticals divide into two structural different families based on
their luminosity profiles at scales greater than 2 kpc (i.e., this is not the
well-known core versus cusp division of ellipticals, Kormendy \etal 2009).  While it
is known that ellipticals divide into two types by isophote shape (boxy versus disky)
and kinematics (supported by rotation and anisotropic velocity distributions),
neither of these physical characteristics are related to the two families by
structure.

\section{Fitting Functions and Isophotal Properties}

The traditional way of understanding galaxy surface brightness profiles is to use
fitting functions.  The most popular of these curves is the de Vaucouleurs or
$r^{1/4}$ fit which uses two parameters (effective radius, $r_e$ and effective
surface brightness, $\mu_e$) and fits a straight line to the points in $r^{1/4}$
space.  The advantage of the $r^{1/4}$ law is that since only two parameters describe
the entire galaxy, and these parameters are correlated by the Kormendy relation, this
means that ellipticals are self-similar as a function of luminosity (i.e., they have
structural homology).  This type of homologous structure is predicted by various
models that use violent relaxation during galaxy formation (Hjorth \& Madsen 1993).

The next generation of fitting functions included the S\'{e}rsic $r^{1/n}$ fitting
function, a modified $r^{1/4}$ law that adds one more parameter, a changing profile
slope $n$.  This suggests a changing curvature to the profile which is not captured
by a single power-law and values for the profile shape can vary greatly depending on
how many and which data points are used to compute the fit.  The curvature index $n$
adds another degree of freedom, allowing for less error in the fits, but does not
provide any additional information as to why elliptical surface brightness profiles
are shaped as they are.  There are well known, and systematic, deviations, from the
$r^{1/4}$ shape with luminosity, and thus the S\'{e}rsic $r^{1/n}$ function does an
admirable job of fitting a profile shape in either the interiors or the outer
envelopes, but not both at the same time (Schombert 2013).

Despite the difficulties with fitting functions, it is well known that galaxy
structure closely follows galaxy luminosity (Schombert 1986).  This is not simply a
statement that galaxy size increases with galaxy luminosity (a proxy for stellar
mass).  It has been shown that any characteristic radius is a smooth function of a
characteristic luminosity or surface brightness.  For example, very early studies
that used an isophotal radius (such as the Holmberg radius or the half-light radius,
Strom \& Strom 1978, Kormendy 1980) found fairly good uniformity over limited ranges
in magnitude ($L \propto r^{1.7}$ for Strom \& Strom, $L \propto r^{0.7}$ for Kormendy).
Technical difficulties in finding standard measures of luminosity and structure
inhibited direct comparison between samples.  Total luminosity is operational simply,
the magnitude where the curve of growth flattens, but total radius is nearly
impossible to define as the there is no sharp edge to a galaxy's gravitational
potential and is undefined by curves of growth. 

A characteristic radius becomes the most difficult structural parameter to determine.
Attempts to determine a radius that encompasses a large percentage of the total
luminosities suffers from large uncertainties due to the high photometric errors at
faint surface brightness levels.  Using a lower percentage of the total luminosity
runs the risk of missing large changes in structure outside the assigned luminosity
level (i.e., half light radius).  Despite these difficulties, refinement of
characteristic radius versus luminosity (Schombert 1987, Graham \& Guzm{\'a}n 2004)
finds a relationship that is well defined within the photometric errors.  The
empirical evidence suggests a testable hypothesis that there exists a unique set of
isophotal properties that define the structure of a galaxy for a given luminosity.
This implies the shape of a galaxy's surface brightness profile is a smooth function
of luminosity as well, which is what is being captured by fitting functions and their
parameters.

To test this idea, various isophotal radii can be compared to search for a
correlation that is greater than predicted by photometric error scatter.  We have
selected the 2MASS elliptical sample (428 objects) from Schombert \& Smith (2012), a
sample of galaxies classified as elliptical in Revised Shapley-Ames catalog (RSA) and
the Uppsala Galaxy Catalog (UGC).  The sample contains all galaxies greater than 5
arcmins in radius but without nearby companions or bright stars.  Distances and
extinction are taken from the NED database.  It is important to note that all the
galaxies in the sample were selected by morphology.  They display no visual evidence
of a disk or SO-like appearance under varying contrasts.  They cover a range of axial
ratio, there is no bias to only observe round ellipticals.  A classical morphologist
would have assigned them strictly as E0 to E6 on the Hubble/Sandage scheme.

\begin{figure}[!ht]
\centering
\includegraphics[scale=0.75,angle=0]{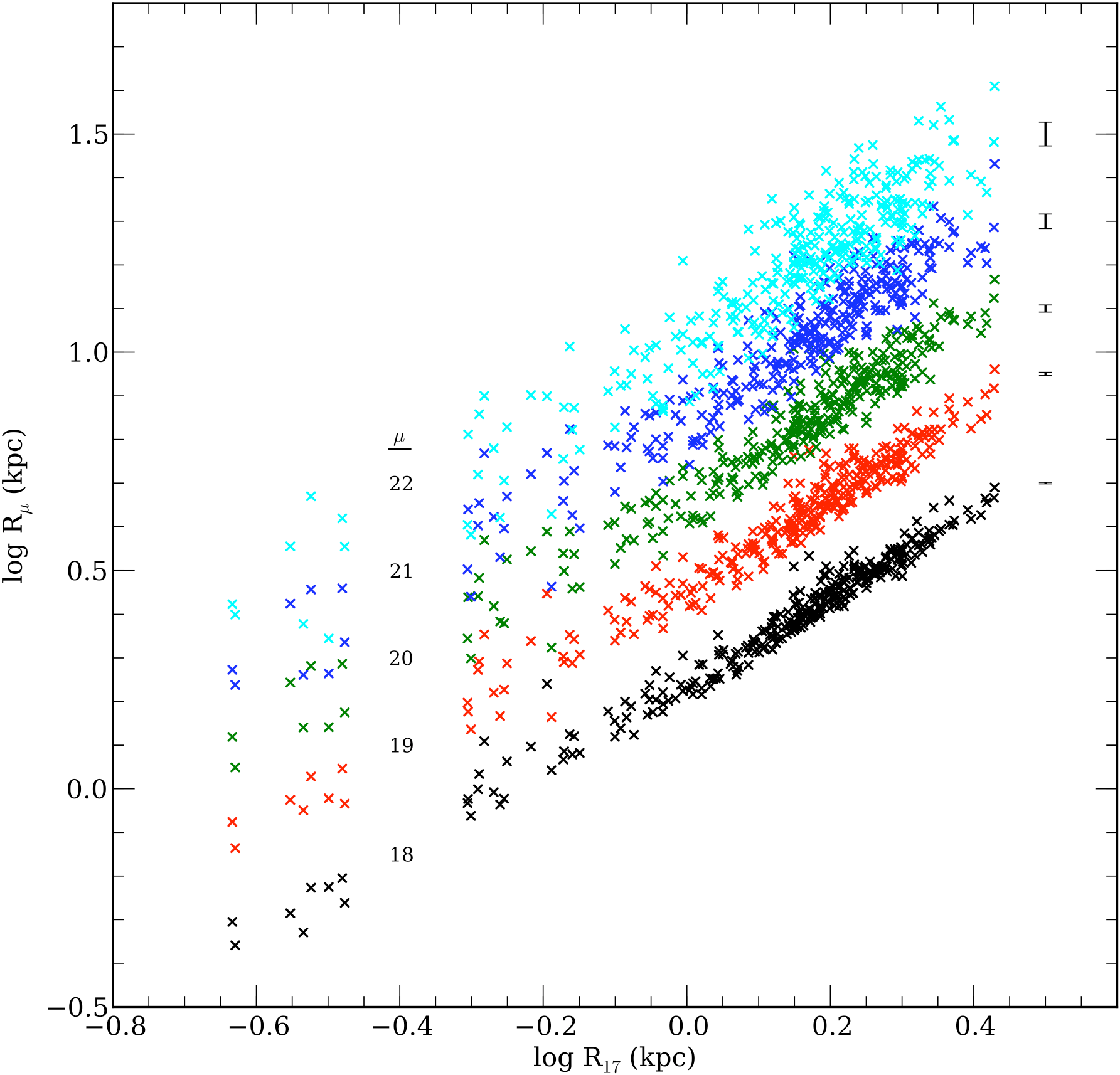}
\caption{\small The isophotal radius relations for 310 ellipticals from the Schombert
\& Smith 2MASS sample.  Each radius is the generalized radius ($\sqrt{ab}$) in kpc at
the isophotal value shown.  Deviations from a linear fit were within the photometric
errors for the brighter isophotes; however, the relationships become distinctly
non-linear at fainter isophotes.  The tight, and nearly linear, relationships is an
argument for structural homology in ellipticals.
}
\label{radii}
\end{figure}

Figure \ref{radii} displays the isophotal radius at the 17 $J$ mag arcsec$^{-2}$
versus the radius at 18 through 22 (the typical $B-J$ color of an elliptical is 4,
for mental conversion to standard $B$ surface brightness values).  These are
generalized radii (for major and minor axis $a$ and $b$, the generalized radius is
defined as $\sqrt{ab}$) in order to normalize mean ellipticity.  The range in total
$J$ luminosity from log $R_{17}$ of $-$0.6 to 0.4 is $-$20.5 and $-$25.0.  Only 310
ellipticals were used for these diagrams, the reasoning will be stated in the next
section.  The trend is, unsurprisingly, for increasing radii at all surface
brightness levels.  The deviation from a linear fit is within the photometric errors
for a majority of the galaxies, although the relationships appear to be slightly
non-linear at higher radii (see below).  

Immediately apparent from Figure \ref{radii} is that ellipticals are remarkably
uniform in terms of structure.  Characteristic radius is strongly correlated with
luminosity (although not linear, see Graham 2005).  Therefore, since each isophotal
radius is also strongly correlated with every other isophotal radius, then the shape
of a ellipticals surface brightness profile is also unique to each luminosity.
Despite the difficulties in deriving structural parameters from fitting functions
(mostly a problem of profile shape), the isophotal size of ellipticals, at all
isophotal levels, is a single function of luminosity.

The homogeneity of ellipticals with respect to structure is a well known fact.
Although their scatter is higher than photometric error in fitting function relations
(Graham 2002), the structural axis of the Fundamental Plane displays the least
variation (Cappellari \etal 2013).  Schombert (1986) displayed both graphically (his
Figure 8) and empirically that ellipticals display a smooth change in structure as a
function of luminosity.  However, the change in shape and profile slope are not
parameterized by fitting functions which, by their mathematical nature, smooth over
irregularities profile as shape not well described by each function.

\begin{figure}[!ht]
\centering
\includegraphics[scale=0.75,angle=0]{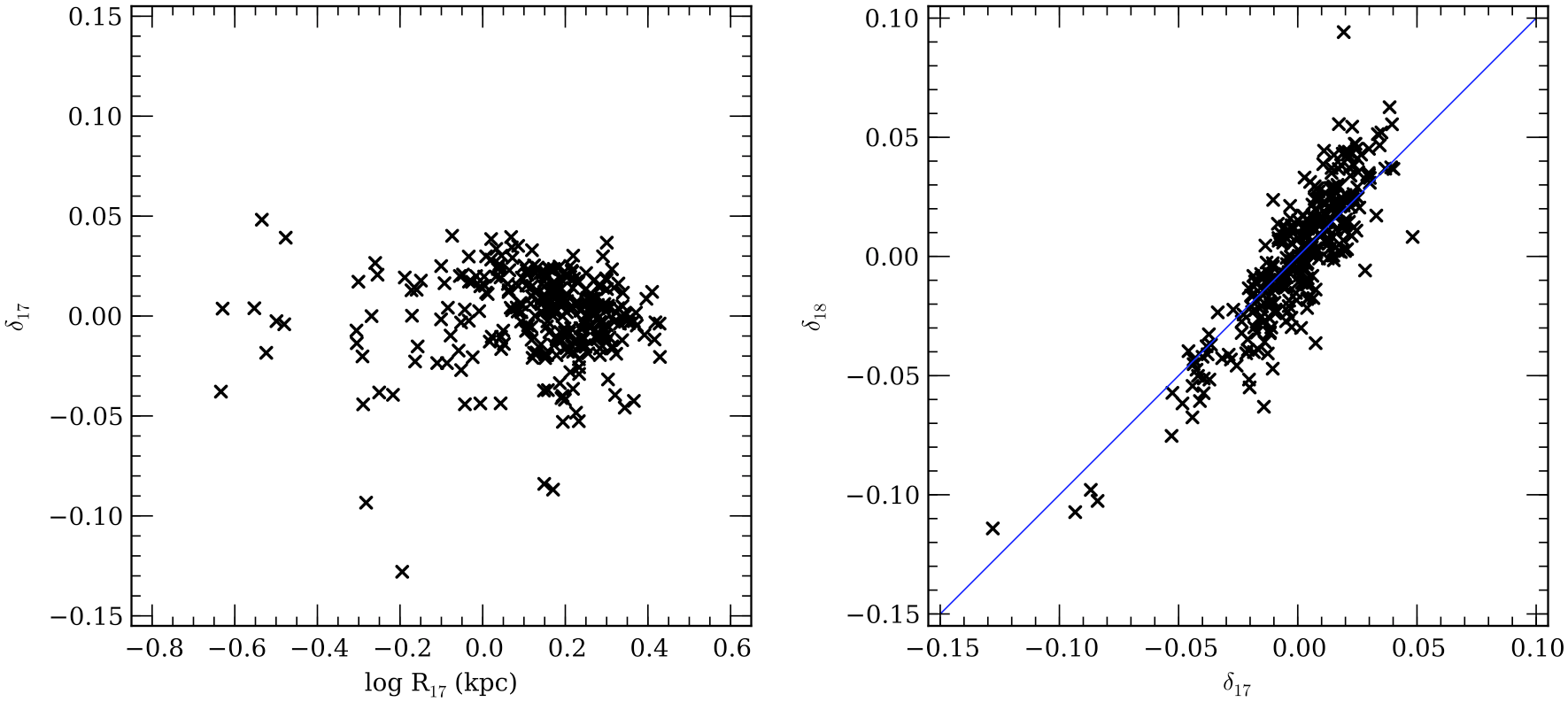}
\caption{\small The left panel displays the residuals from a linear fit between
$R_{17}$ and $R_{18}$ from Figure 1.  There is a weak negative slope indicating that 
the relationship may be slightly non-linear.  The right panel displays the
correlation between the residuals in $R_{17}$ and $R_{18}$.  While the correlation
does exist, the lack of a one-to-one slope confirms the indication from the left
panel of non-linearity to the isophotal radii relationships.  To form template
profiles, a more sophisticated technique will be needed other than simple linear
fitting.
}
\label{residuals_iso}
\end{figure}

A test for non-linearity is found in the left panel in Figure \ref{residuals_iso}
which displays the residuals from a straight line as a function of $R_{17}$.  There is a
indication of residuals being slightly more negative at larger radii, suggesting
curvature to the empirical relation between the isophotal radii.  This effect is
larger at fainter isophotes.  The right panel in Figure \ref{residuals_iso} display
this change by comparing the residuals from a linear fit between $R_{17}$ and
$R_{18}$.  A clear difference from a one-to-one relationship is evident.  This means
that characterizing the shape of elliptical profiles will need to be more
sophisticated than simple straight line fits to the isophotal radii relationships.

\section{Template Construction}

If the shape of an ellipticals surface brightness profile is also a function of
luminosity (as suggested by Figure \ref{radii}), then it should be possible to
generate a series of generalized or template profiles from the isophotal relations or
mean averaging the profiles themselves.  We follow the prescription of Schombert
(1986) where mean templates were generated from photographic $V$ surface photometry
of cluster ellipticals.  The motivation for that study was simply to find an
empirical profile shape in order to subtract the underlying galaxy from cD envelopes
to estimate their luminosities (Schombert 1988), but the basic technique is the same.
The conditions found in $V$ from Schombert (1986) are identical to the photometric
relationships found for the 2MASS $J$ sample, such that the isophotal radii
relationships were nearly linear and their scatter was purely photometric.

The $J$ profiles from the 2MASS sample have similar depth as the original $V$ study,
with similar plate scale (i.e., resolution) but using digital devices with less RMS
noise around each elliptical isophote and better sky subtraction (although the sky at
$J$ is a factor of three times brighter than the sky at $V$).  To construct the
templates from the $J$ profile, we have selected a statistical framework that uses
the middle of the profiles first, then weighting the outer and inner data points by
photometric error.  While the fitting was automated, visual inspection is made of the
initial templates to avoid systematics that mimic real variations in structure.

For the first pass on creating the templates, the galaxies were grouped by luminosity
at the radius of 4 kpc.  This radius was chosen because it is a distance outside PSF
effects from the core but at sufficiently high surface brightnesses that error due to
photometric and sky uncertainties are small.  The galaxies for each group had
generally the same luminosity and the average of their luminosity predictably grew
for each successive grouping.  This first stage of template construction proved to be
unstable for two reasons; 1) an inner aperture magnitude was slightly sensitive to
the core versus cusp behavior found in many ellipticals (Kormendy \etal 2009) and 2)
a significant number of the ellipticals deviated from the average values (see below).

The second pass for template construction used isophotal radii as the normalizing
metric.  This proved to be more stable as an aperture magnitude is an integrated
quantity, whereas a characteristic scale length is only sensitive to photometric
errors at that particular radii.  For the templates, each galaxy is averaged at the
same radial point using linear interpolation between surface brightness levels.  Once
the group has been averaged, the aperture magnitude is numerically determined from
the artificial profile as a crude identifier (i.e., the templates are labeled by
their 16 kpc aperture magnitudes).

As is the case in a heterogeneous sample, some galaxies are a better fit to the
average than other galaxies in the luminosity groups.  And, more importantly, there
will be some slight variation due to the range in luminosity within each group. To
account for this, we weighted the galaxies by their standard deviation, $\sigma$, from
the average and re-averaged the galaxies luminosity profiles using the inverse of
$\sigma$; $\sigma=\sqrt{\sum_i\frac{1}{N-1}(\mu_i-\bar{\mu})^2}$, where $\bar{\mu}$
is the averaged value of the surface brightness at that radius and $i$ spans all of
the galaxies for the average.  Once the templates were assembled, they were smoothed
by a fourth order spline.  This was done mostly to minimize photometric errors at
larger radii where data is less reliable, but also to eliminate an artificial jagged
appearance to the templates due to forced selection of radii bins.  The smoothing was
never greater than 0.005 mags arcsecs$^{-2}$.

\begin{figure}[!ht]
\centering
\includegraphics[scale=0.9,angle=0]{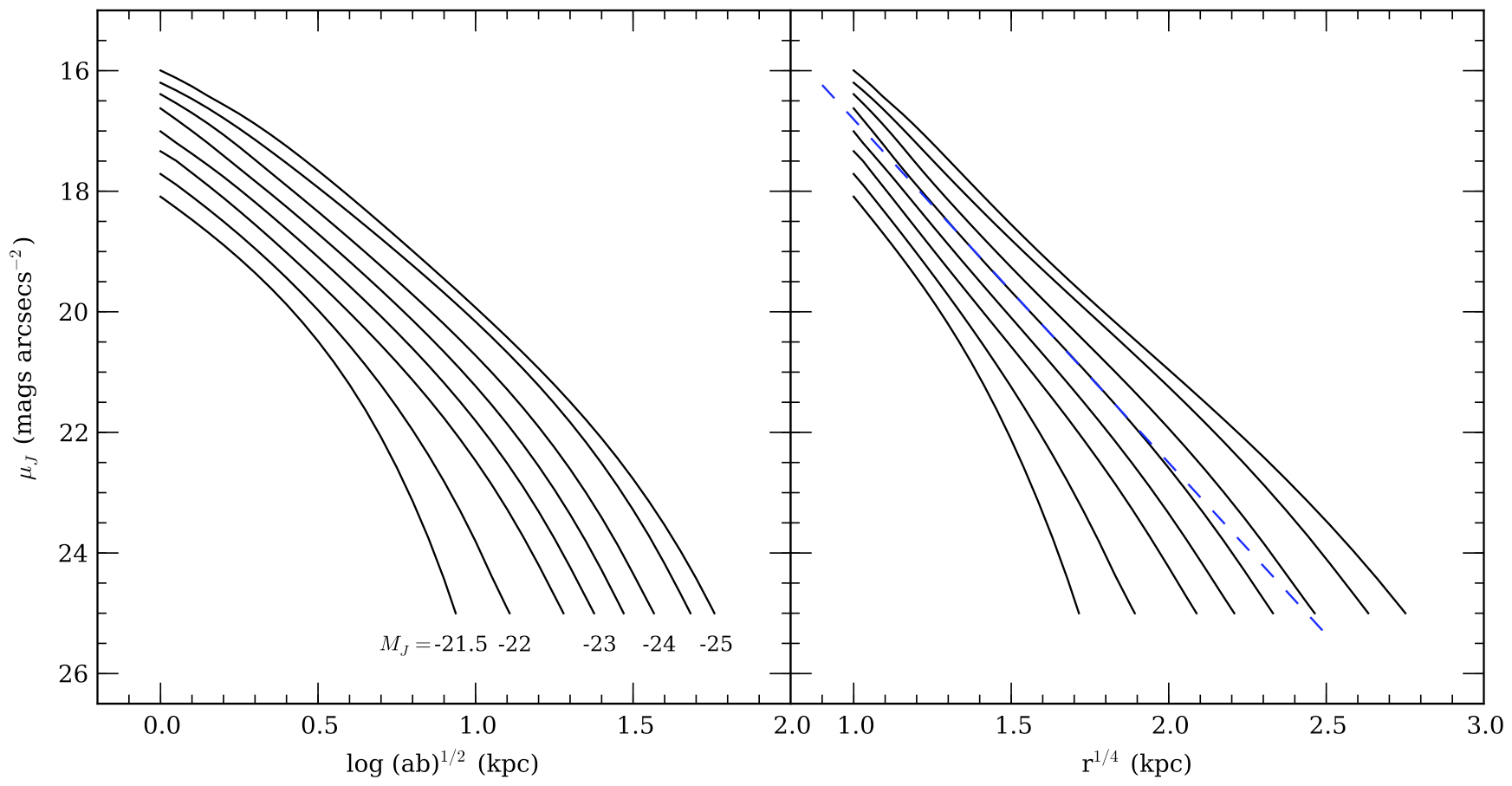}
\caption{\small Generalized template profiles from 2MASS $J$ surface photometry
constructed from 308 morphologically pure ellipticals.
Plotted as log generalized radius ($ab^{1/2}$) versus surface brightness on the left
and r$^{1/4}$ on the right (for visual comparison to the de Vaucouleurs relationship).
The templates are generated from a look-up table at fixed isophotal radius, but are
parameterized by their 16 kpc aperture magnitude (shown along the bottom).  From the
raw profiles, 95\% of the galaxies lie between $-$21.5 and $-$25, although slight
extrapolation beyond these limits are reasonable for comparison to brightest cluster
galaxies (i.e., D and cD) and dwarf ellipticals.  The blue dotted line is a least
square $r^{1/4}$ fit to the $-$23.5 profile as a demonstration to the limitations of
the $r^{1/4}$ law.  The templates demonstrate that the structure of ellipticals is a
uniform function of luminosity.
}
\label{templates}
\end{figure}

The resulting profiles are shown in Figure \ref{templates}.  The actual template
construction is such that generalized shapes at various radii are maintained in a
lookup table of twenty templates.  Any particular template, as shown in Figure
\ref{templates} is interpolated from the table and output with a 16 kpc magnitude as
an identifier.  The characteristics of the profiles are very similar to the templates
constructed in $V$ (Schombert 1986) in the sense of a smooth logarithmic shape with a
distinct dropoff at large radii.  The dropoff at large radii is an important point,
for the outer slope is such that all galaxies integrate to a finite luminosity (this
is not the case for many cD galaxies).  In other words, all the outer profile shapes
are less than $L \propto r^{-2}$ explaining why curve of growth measurements for
total luminosity in ellipticals frequently converge.  The inner and midsection slopes
of the profiles decreases gradually with increasing luminosity and the outer dropoff
becomes steeper for fainter galaxies.

The right panel of Figure \ref{templates} displays the same templates plotted in
$r^{1/4}$ space (where the $r^{1/4}$ law is a straight line).  The early adoption of
the $r^{1/4}$ is understandable based on visual inspection that profiles are
$r^{1/4}$ in shape in the mid regions for the brighter ellipticals.  The $r^{1/4}$
shape clearly fails for the outer isophotes at all luminosities, and fails for all
profiles with luminosities less than $-$22.  The advantage of the S\'{e}rsic
$r^{1/n}$ function is that the $n$ parameter captures this outer envelope curvature
(or inner profile flatness depending on the region fitted).

The procedure for fitting a particular galaxy's surface brightness profile to a
template depends of the type and quality of the data.  The algorithms that fit
ellipses to 2D images tend to have more ellipses at smaller radii due to the higher
S/N for those isophotes.  Ellipses at larger radii tend be more widely spaced so as
to decrease the photometric noise by using more pixels.  Thus, template fitting was
done in log radius space so that the isophotal radii are evenly spaced.   Both inner
and outer isophotes are weighted less, the outer isophotes by photometric error and
the inner isophotes by their distance from the seeing correction region.

\begin{figure}[!ht]
\centering
\includegraphics[scale=0.90,angle=0]{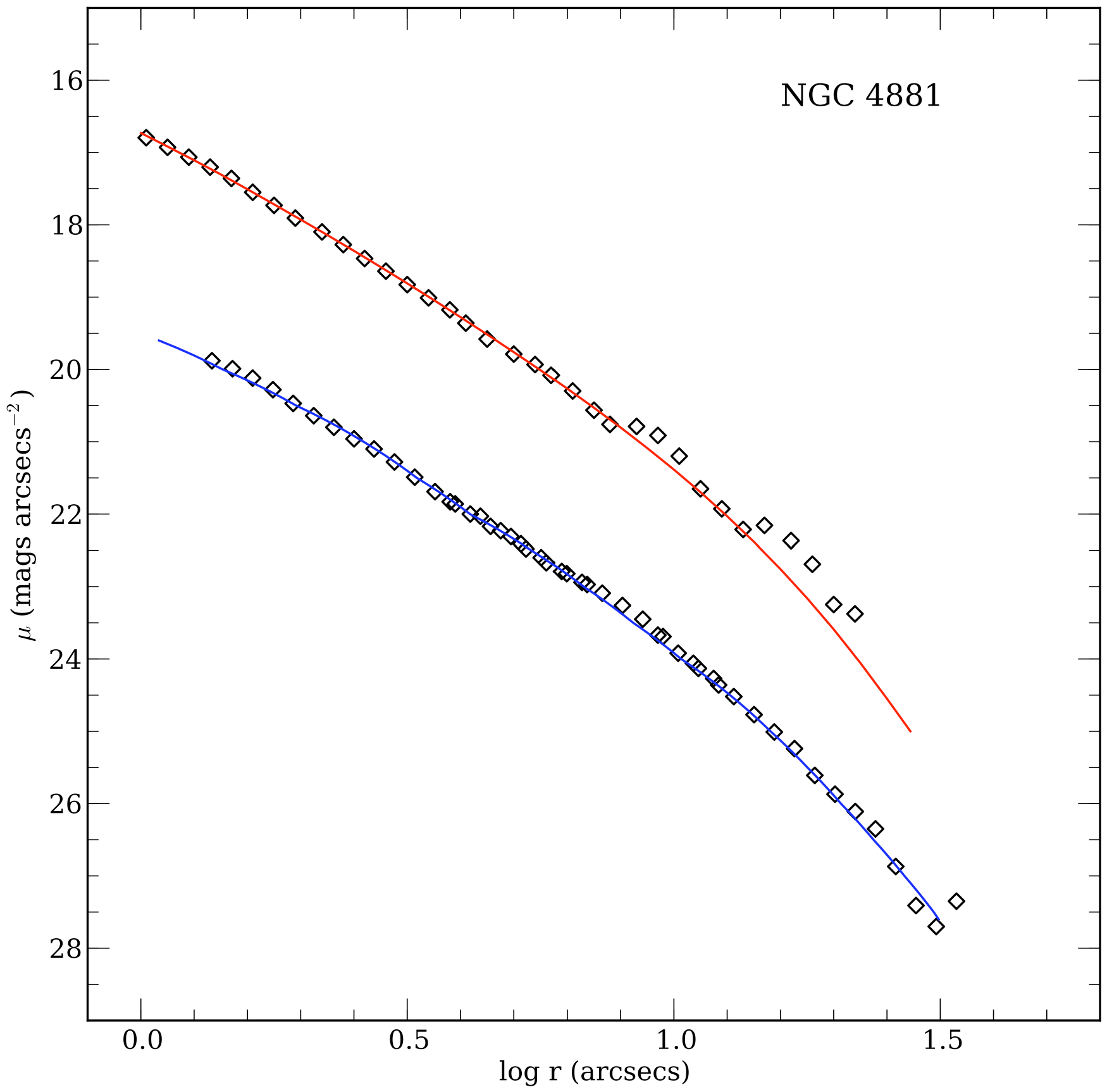}
\caption{\small A comparison of the template fits for NGC 4881 (an E0 in the Coma
cluster) in Johnson $V$ (blue) and $J$ (red).  The $V$ data is from Schombert (1986), old
photographic and CCD images. The $J$ is from 2MASS, the mean $V-J$ color for NGC 4881
is 2.70.  The template correspondence is
excellent considering the difference in image material and time.  The difference in
slope between the $V$ and $J$ data is due to a color gradient of 
$\Delta(V-J)/{\rm log\ r} = 0.15$.  
}
\label{ngc4881}
\end{figure}

An example of a good fit is shown in Figure \ref{ngc4881}, a comparison of the $V$ and
$J$ surface brightness profiles and templates for NGC 4881.  NGC 4881 is located in
the Coma cluster and is a common test galaxy for surface photometry due to its nearly
perfectly circular shape and isolation from other galaxies and bright stars.  Shown
in Figure \ref{ngc4881} is a comparison between Johnson $V$ templates from Schombert
(1986) and the current set of templates for 2MASS $J$.  The original $V$ templates
were based on photographic and early CCD imaging, but display the same slope and
scaling relations as the newer $J$ templates (when color gradients are taken into
consideration).

The NGC 4881 data displays a color gradient of $\Delta(V-J)/{\rm log\ r} = 0.15$,
which is near the mean value for bright ellipticals (La Barbera \etal 2012).  Thus,
templates are slightly dependent on the filter passband used in making the surface
brightness profiles and color gradients make comparison problematic, although
gradients tend to be systematic with luminosity (Roig, Blanton \& Yan 2015).
Ideally, one would covert the surface brightness profiles at $J$ into mass density
using the information from color combined with stellar population models to derive a
correct $M/L$ with surface brightness (McGaugh \& Schombert 2015).

\section{Structural Homology}

One of the earliest goals of studying galaxy luminosity profiles was to investigate
whether the structure of ellipticals was self-similar, meaning that a more massive
galaxy was, simply, a scaled up version of a less massive galaxy with a uniformly
larger scalelength and more energetic internal kinematics.  Early evidence on the
$r^{1/4}$ nature of elliptical profiles, and the linearity in scaling relations
derived from $r^{1/4}$ fits, supported this view of structural homology (Prugniel \&
Simien 1997).  However, more detailed studies of velocity dispersion, luminosity and
scalelength (the core components of the Fundamental Plane) revealed non-homology
(Prugniel \& Simien 1996).  Non-homology was suspected to be mostly due to deviations
from stellar population effects, but could also have their origin in deviations from
the $r^{1/4}$ law.

Finding uniformity in structure is also a goal for investigating the origin of
structure by formation processes.  For example, early numerical simulations of
dissipationless collapse demonstrated that self-gravitating systems can interact and
relax to reach a universal structure despite varying initial conditions (van Albada
1982; Miller 1988; Barnes 1989) leading to the hope that a statistical
mechanical theory of galaxy formation could be obtained (see Hjorth \& Madsen 1991).
Additional modifications to the models, using a finite escape energy, found that
deviations from the $r^{1/4}$ shape would follow naturally from a pure dissipationless
scenario.  The various configurations would result in a galaxy with an anisotropic
velocity distribution in a triaxial shape, similar to what is observed for bright
ellipticals but with insufficient flattening to describe faint ellipticals.

Complications arose from a pure violent relaxation interpretation.  For example, high
central densities for low luminosity ellipticals are difficult to reproduce by
dissipationless processes and suggest some dissipative component to galaxy formation
is required.  The existence of stellar population gradients (particular in
metallicity) are also difficult to reproduce without dissipation.  However, more
sophisticated models were able to produce $r^{1/4}$ profiles that mimic real
galaxy distributions with characteristic deviations at various radii.  In particular,
they predicted the dropoff in surface brightness below the $r^{1/4}$ law at large
radii and lower central densities than predicated by the $r^{1/4}$ law (see Figure 3
in Hjorth \& Madsen 1991 and Figure 2 in Hjorth \& Madsen 1995).

The templates from the last section do demonstrate that ellipticals are homologous
with respect to structure in a limited sense (so-called weak homology).  A majority
of ellipticals do have structure that varies uniformly with luminosity (i.e., a
particular surface brightness profile is identified at every stellar mass).  However,
the template profiles do not vary in a linear fashion with respect to scalelength nor
characteristic surface brightness (i.e., luminosity density).  And each profile is
not self-similar to any fainter or brighter profile, thus the difficulty that fitting
have in functions reproducing elliptical structure as a function of luminosity
(Schombert 2013).  A shape parameter (equivalent to the S\'{e}rsic $n$ variable) must
be added to describe the templates.  This introduction of a parameter that is not
self-similar destroys absolute homology, although the deviations are small.  Thus,
homology is a close approximation to the range of elliptical structure (and not a
major contributor to the tilt in the Fundamental Plane Prugniel \& Simien 1997), but
ellipticals as a class have structural features that vary with galaxy mass and are,
by definition, non-homologous (Graham \& Colless 1997).

Templates, while more accurately describing structure and change in structure,
complicate the interpretation of structure as presented by fitting functions and
comparison to theoretical predictions.  For example, it is relatively easy to fit
density results from N-body simulations and compare the scale lengths with fitting
function results (Burkert 1993).  However, it would be much more complicated to
convert those mass density profiles to luminosity density (with the stellar
population uncertainties) and compare that to our templates.  The templates recover
all the known surface brightness relations (such as effective radius, $r_e$, versus
effective surface brightness, $\mu_e$).  Table 1 displays the best fits to the five
templates found in Figure \ref{templates} to the $r^{1/4}$ and the S\'{e}rsic
$r^{1/n}$ fitting functions.  The Table is divided into three parts outlining the
fits under the conditions of described in Schombert (2013) for inner fits (inside the
empirical half-light radius, $r_h$), outer fits (outside 80\% of $r_h$) and fits for
the full profile.  Although Table 1 followed the technique outlined in Schombert
(2013) for profile fitting, these fits do not capture the diverse range in fitting
parameters.  For example, the S\'{e}rsic $n$ parameter is typically lower for the
templates than the mean value for $n$ from actual data (see Figure 5 in Schombert
2013).  This is due to the fact that the template profiles have no photometric error
assigned to their values (although, in theory, one could assign an uncertainty value
based on the dispersion in the isophotal relations).  Thus, the fitting routines give
equal weight over the range in surface brightness of the template.  This results in
more curvature at fainter surface brightness levels then actual data with larger
photometric errors at larger radius and Table 1 is presented for reference solely to
the shape of the templates.

\begin{figure}[!ht]
\centering
\includegraphics[scale=0.90,angle=0]{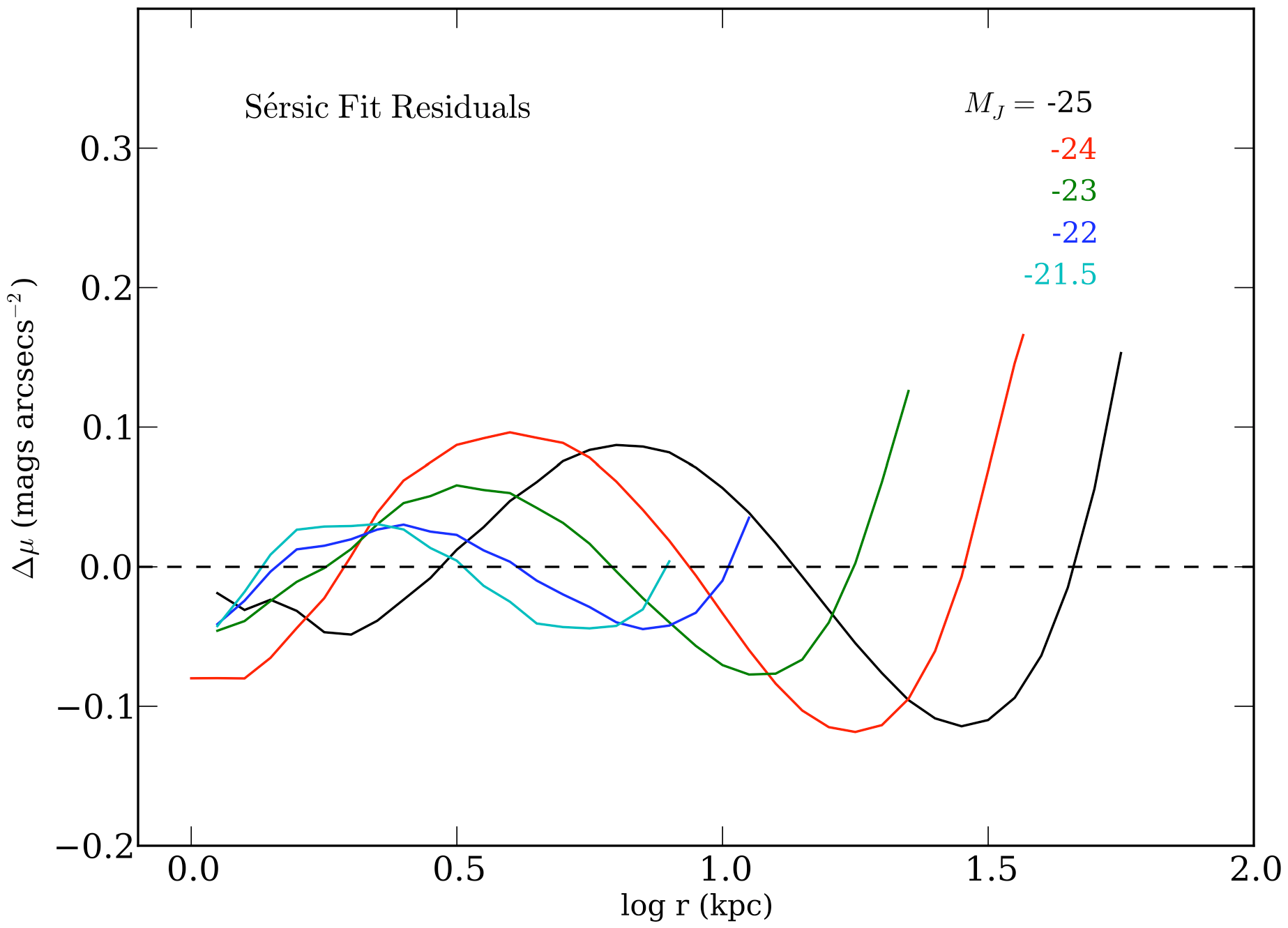}
\caption{\small The surface brightness residuals for the five templates in Figure
\ref{templates} versus log radius.  Each luminosity bin is shown as a separate color.
While the residuals are on average small, indicated that the S\'{e}rsic $r^{1/n}$
function is a adequate describer of an ellipticals profile, there are systematics
uniform by luminosity that state that ultimately the surface brightness profiles are
not S\'{e}rsic $r^{1/n}$ shape.  We note that the residuals have many of the features
predicted by violent relaxation simulations (Hjorth \& Madsen 1995).
}
\label{fit_residuals}
\end{figure}

The residuals to the full S\'{e}rsic function fits are shown in Figure
\ref{fit_residuals}.  The limitations to the S\'{e}rsic $r^{1/n}$ function are
outlined in Schombert (2013), although it is the best "French curve" fitting function
available.  The resulting fit parameters to the templates reproduce all the known
scaling relations from the S\'{e}rsic function, including the Photometric Plane
(Graham 2002).  However, Figure \ref{fit_residuals} displays systematic residuals
that are consistent from each luminosity bin.  Thus, while an adequate describer of a
typical elliptical profile, the surface brightness profiles of ellipticals are
ultimately neither $r^{1/4}$ nor S\'{e}rsic $r^{1/n}$ in shape.  In particular, the
upturn in residuals at large radii will result in statistically higher $n$ indices in
observed profiles with real photometric errors.  

We also note that the residuals in Figure \ref{fit_residuals} can be reproduced by
statistical mechanical violent relaxation models (Hjorth \& Madsen 1995).  Their
Figure 2 displays many of the same features as in Figure \ref{fit_residuals}, such as
an extended envelope for bright ellipticals (for galaxies with deep central
potentials), a depressed envelope for faint ellipticals (with shallow central
potentials) and fainter core region for all luminosities.  The difficulty in
interpretation is that the central potential parameter that defines the models has a
large range of values unconfined by observations.  While it is encouraging that
similar profile shapes can be produced by simple dissipationless scenarios, this is
inadequate as a full galaxy formation theory.

The residuals do indicate some subsistence to the technique used by Huang \etal
(2013) where three components are fit to an elliptical profile; a core ($r < 1$ kpc),
an intermediate region ($r \approx 2-3$ kpc) and an outer envelope ($r > 10$ kpc).
From Figure \ref{templates}, we can see that the profiles divide into the same three
regions, a core (not well sampled in the 2MASS images), a $r^{1/4}$ middle region and
an outer envelope that either extends above the $r^{1/4}$ shape (at high
luminosities) or below (at faint luminosities).  However, without a physical basis
for this division in structure, the multi-component technique is simply an elaborate
spline curve to the data and it is not surprising that a three component model is a
better match to elliptical structure as displayed by the templates in Figure
\ref{templates}.  Whether it contains any underlying structural information is
unknown.

\section{Two Families of Ellipticals}

\subsection{Normal versus D Ellipticals}

During the initial template construction, using 468 elliptical profiles, the averaged
profiles failed to converge (numerically) to smooth templates with scatter less than
the photometric errors.  Inspection of the residuals between the actual galaxy
profiles and averaged templates revealed that the problem was due to a specific
subset of the profiles with consistently different shapes per luminosity bin than
most other elliptical profiles.  In particular, a plot of template residuals versus
radius displayed a `cross' pattern where 2/3rd's of the galaxies formed one leg with
a negative slope and 1/3rd formed the second leg of positive slope.

\begin{figure}[!ht]
\centering
\includegraphics[scale=0.75,angle=0]{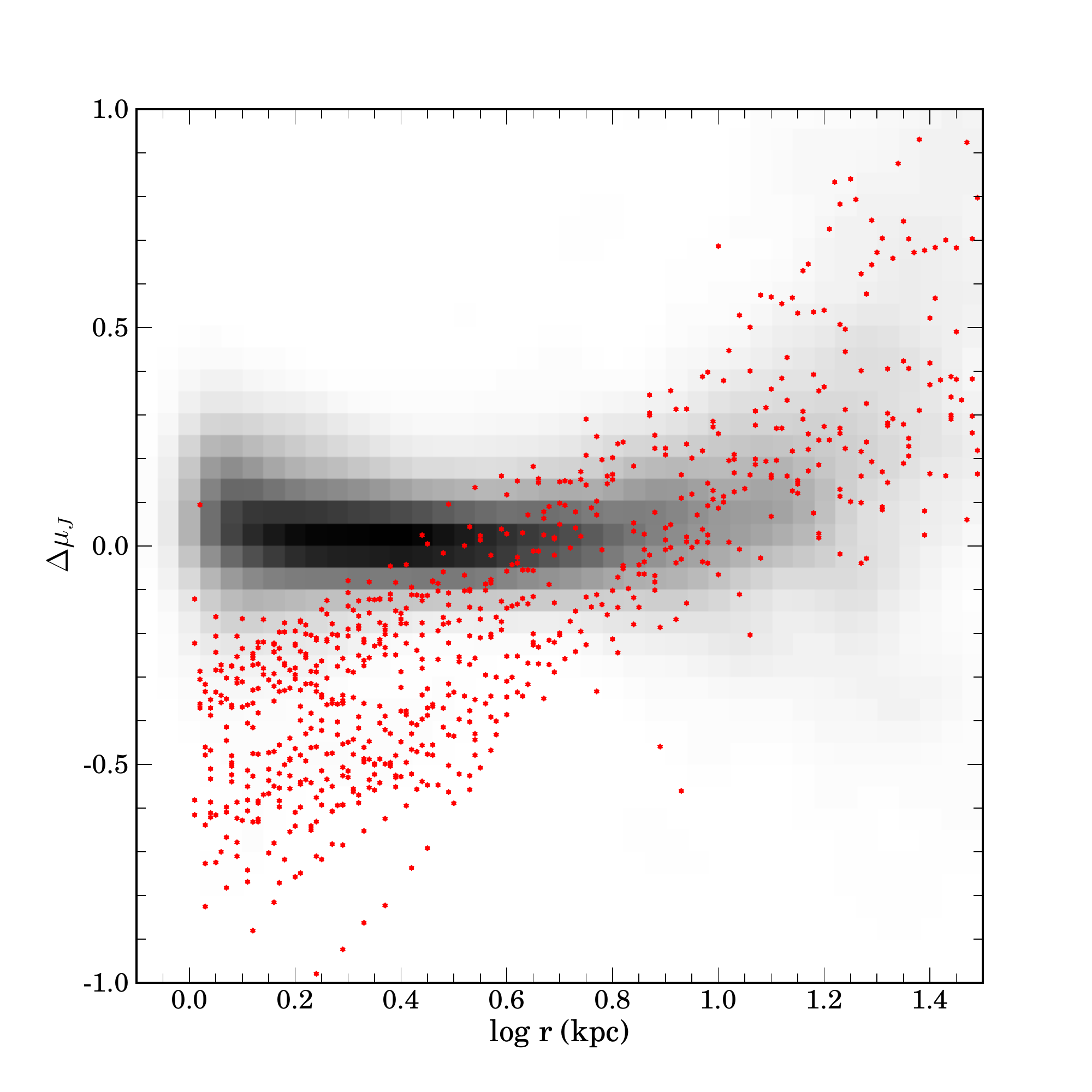}
\caption{\small The difference between the surface brightness profiles of 311 ellipticals
and mean template profiles are shown as the greyscale density image.  Over
90\% of the surface brightness points are with 0.15 mags of the templates.
However, for many galaxies (117 morphologically classed as ellipticals), their
surface brightness profiles deviate in a systematic fashion from the
templates (i.e., not a poor fit, a different shape).  A subset of the deviant
profiles are shown as red symbols.
}
\label{template_residuals}
\end{figure}

Fitting only the first type resulted in a convergence on a set of templates that was
well matched to 2/3rd's of the sample.  Figure \ref{template_residuals} displays the
residuals from this second fit, where the greyscale are the difference between the
surface brightness profiles and the final templates ($\Delta\mu_J$) displayed as a
Hess density plot.  The difference between the templates and first type of data were
less than 0.15 mag arcsecs$^{-2}$ for 90\% of the subset.  Many of the second set of
galaxies (45 of them shown as red symbols in Figure \ref{template_residuals}) clearly
deviate in a systematic fashion from the templates.  Through an iterative procedure,
we eliminated a majority of the second type of profiles from the sample and
calculated templates using only profiles from the first type.  These final templates
are the ones shown in Figure \ref{templates}.

Ultimately, 157 (33\%) profiles were identified to deviate in a specific fashion from
the templates (and were rejected from template construction).  Some profiles were
simply irregular and may be the result of recent interactions which has disturbed the
luminosity distribution (40 of the 157).  However, a majority of the deviant profiles
(117 profiles) are more extended than our first type (larger radii per surface
brightness) but varied with luminosity in the same fashion as the normal ellipticals
(larger and brighter in surface brightness with increasing luminosity).  Some
appeared to have two $r^{1/4}$ shaped components, although at much larger radii than
expected for a bulge+disk S0 classification (Andreon \& Davoust 1997).  Thus, {\it it
is clear from our template analysis that there exist two distinct families of
ellipticals as classed by surface brightness profiles}. 

We emphasize that the two families, as outlined by structure, are not the same as the
core versus cusp structure differences (see Kormendy \etal 2009).  The differences in
structure between the two families is strictly limited to structure well outside the
core regions ($R > 2$ kpc).  In fact, there appears to be no correlation between core
and cusp shaped interiors and the two families exteriors.  We also note that this
division into two families is also unrelated to the proposal by Kormendy \& Bender
(1996) that ellipticals are divided into boxy and disky isophote families (see
\S 5.3).  The surface profile shape is unrelated to isophote shape.

This discovery would be completely missed by studies using fitting functions or even
multi-component fitting functions (e.g., Huang \etal 2013).  For, with a sufficient
number of variables, any shape can be fit and the resulting scaling relations blur
the distinction between the two families.  In fact, the galaxies with the most
prominent third component from profile fitting by Huang \etal mostly fall in our
second class of ellipticals.  The primary distinction between the two types of
profiles is their outer slope.  A slope that is easily mistaken as a larger
S\'{e}rsic $n$ parameter or a slightly larger $r_e$ in $r^{1/4}$ fits.  The two types
of profiles do not separate in any scalelength or surface brightness relationship and
are only discovered by comparison to templates.

\begin{figure}[!ht]
\centering
\includegraphics[scale=0.75,angle=0]{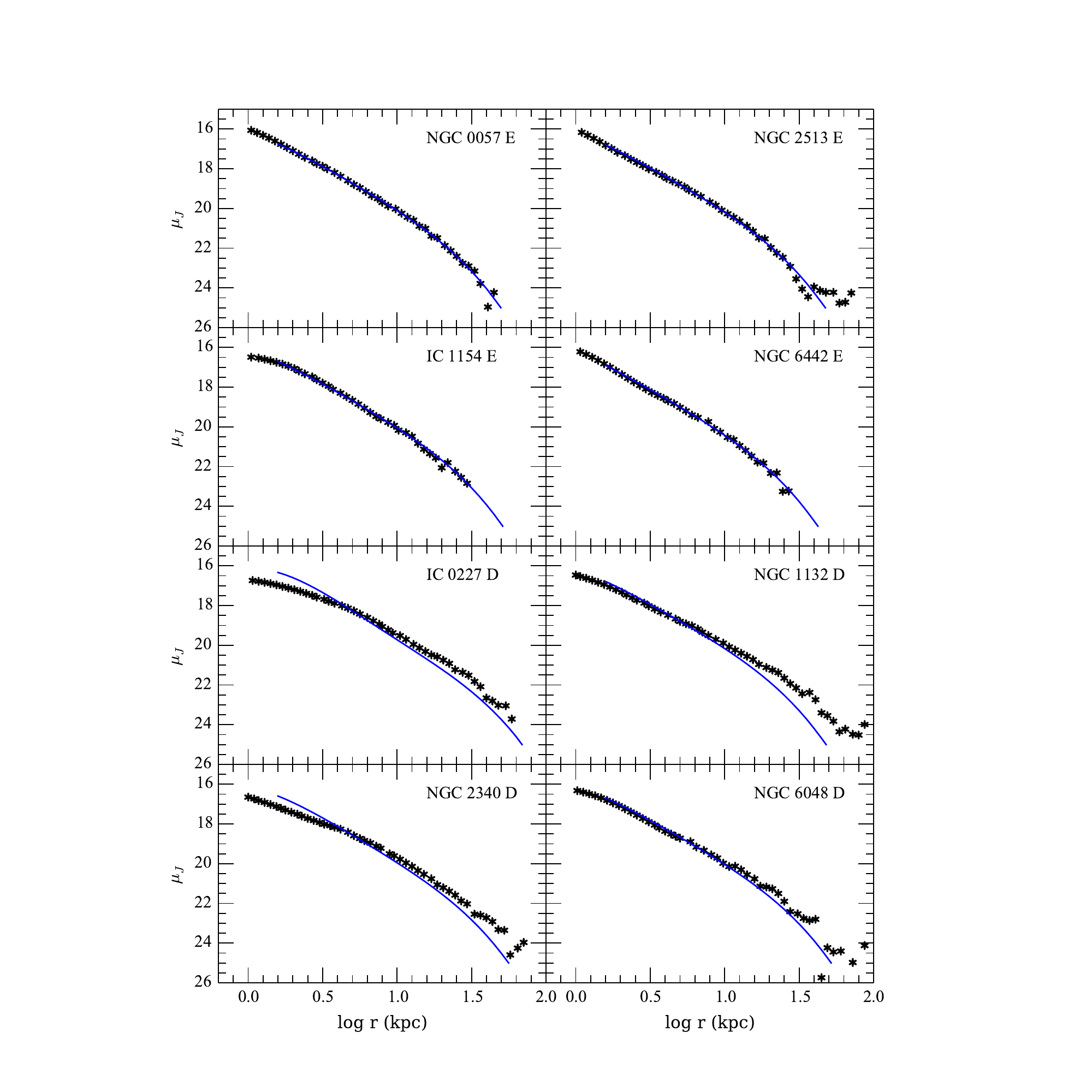}
\caption{\small Examples of template fits to four galaxies with good fits
(classified as ellipticals) and four galaxies with poor fits (classified as
D ellipticals).  The D ellipticals display shallower profile slopes, compared to
templates.  Some, such as IC 0227, have suggestions of two components, but this is
not common for the D ellipticals indicating they are not related to S0's.
}
\label{8panel}
\end{figure}

Examples of the two types of profiles are shown in Figure \ref{8panel}.  This second
type of elliptical is identical in morphology, isophotal characteristics and profile
shape to the first type, but has a shallower slope than other ellipticals of the same
luminosity.  There are many examples of poor template fits based on an irregular profile
shape, but we have reserved the designation of the second family to those profiles which
are under luminous in the interiors and brighter in surface brightness in the outer
envelope (i.e. shallow).  Over luminous profiles produce a `diffuse' appearance on
photographic plates (in the language of surface brightness photometerists).  There
already exists a morphological class for diffuse ellipticals, the D class (Matthew,
Morgan \& Schmidt 1964), so we have designated this second type of elliptical as D
galaxies.  Although Morgan \& Lesh (1965) refined the D class to apply only to the
brightest member of a rich cluster of galaxies (BCM), a study of cluster ellipticals
found several examples of D class ellipticals that were not the 1st, 2nd nor 3rd
ranked in a cluster (Schombert 1986).  For the rest of this paper, we will designate
all galaxies that fit the templates as normal ellipticals, and those which deviate
from the templates with shallower profiles as D ellipticals.  Both normal ellipticals
and D ellipticals have complete rotational symmetry, as defined by the original
Hubble elliptical classification criteria.  Both normal ellipticals and D ellipticals
have smooth surface brightness profiles that decrease uniformly with radius.

Before assigning a new category to the family of ellipticals, we considered the
possibility that the second family with shallower profiles were misclassified S0's.
A large bulge and shallow disk might mimic the second type of profile.  However,
there is no direct connection between the D ellipticals and S0's, for S0 galaxies
have mean ellipticities that are much flatter than the mean for D ellipticals (see \S
5.3), in our study, cover a range in ellipticity (see \S 5.3).  The S0$_1$ class is
the closest in appearance, which only distinguish from ellipticals by the flatter
intensity gradient.  But S0$_1$'s are signaled by a distinct break in the gradient of
their surface brightness profiles that displays a shallower disk region, not a
strictly shallower profile seen for D ellipticals.  While some D ellipticals appear
to have a two component profile shape (e.g., NGC 6048 in Figure \ref{8panel}), they
do not have the characteristic bulge+disk profiles that define the S0 class (i.e.,
the inner component is much larger than a typical bulge).

The D ellipticals are systematically larger than normal ellipticals at any particular
isophotal level and we were unable to construct reliable templates of D elliptical
profiles.  It appears that they are not as consistent in structure as a function of
luminosity as normal ellipticals (which would not be the case if they were misclassified
S0's); however, the number of profiles was less than a quarter of the profiles
available to the construction of elliptical templates and may simply represent small
number statistics.  In the next sections we will explore the properties of D
ellipticals compared to the normal ellipticals in our 2MASS sample.

\subsection{Luminosity and Local Density}

\begin{figure}[!ht]
\centering
\includegraphics[scale=0.80,angle=0]{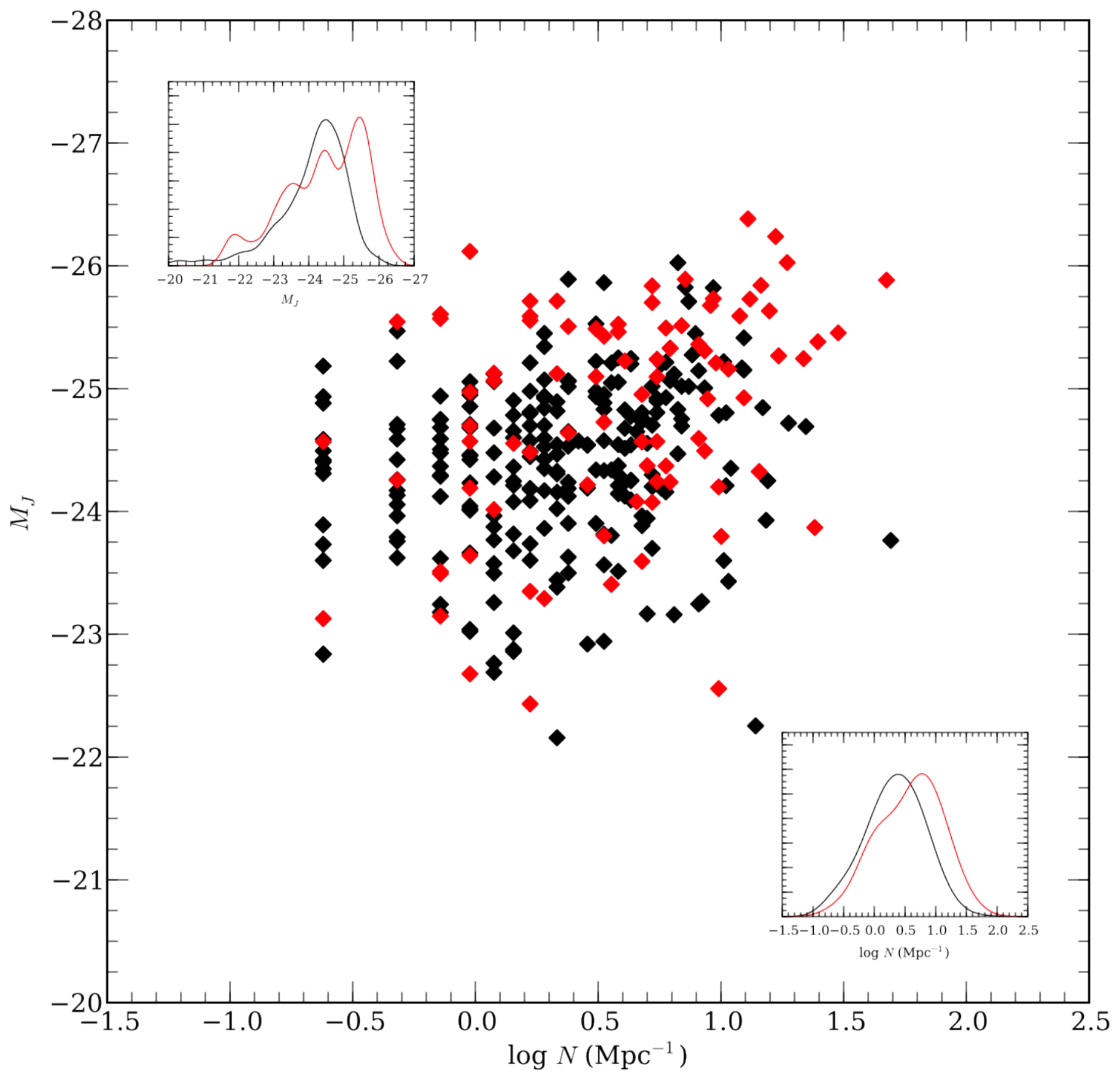}
\caption{\small Total $J$ magnitude versus local density (within 1 Mpc).  Inset in
the upper left and lower right are scaled normalized histograms of the distribution
of total magnitude and local density.  Black symbols are normal ellipticals, red
symbols are D ellipticals.  The total magnitudes are deduced from curves
of growth on the original 2MASS frames.  The histograms are normalized to 0.5 mags
and scaled such that the peak for ellipticals are D ellipticals are identical.  The
local density is determined from the redshift widget in NED.  Their are slightly more
D ellipticals at higher luminosities and local densities, but they also cover the
same range as the normal ellipticals in the sample.
}
\label{mag_comp}
\end{figure}

Figure \ref{mag_comp} displays the luminosity and local density differences between
normal ellipticals and D ellipticals.  The total magnitudes are determined by
asymptotic fits to the curves of growth in the original 2MASS images (see Schombert
\& Smith 2012).  All the galaxies in the sample converged to well determined total
fluxes, ranging from $-$20 to $-$26 $J$ mags.  The average total absolute $J$ mags
for the sample of normal and D ellipticals is identical but, as can been seen from
upper left inset in Figure \ref{mag_comp}, their distribution of luminosities differs
significantly.  There are slightly more D ellipticals at brighter luminosities,
although the D ellipticals cover the same range of luminosities as the ellipticals in
the sample (i.e., there is no deficiency of D ellipticals at any luminosity).

Typically ellipticals divide into two classes by luminosity in plots of total
luminosity versus scalelength (either effective radius, half-light radius or
isophotal radius).  The brighter ellipticals have a slightly different relationship
between luminosity and scalelength ($L \propto r^{0.7}$, see Figure 8 Schombert 1987)
with a break at $M_J > -24$.  The fainter ellipticals display a steeper slope with
luminosity ($L \propto r^{1.6}$).  This has been assumed, in the past, to reflect a
shift in the underlying kinematics for bright ellipticals, which typically have
little rotation, while fainter ellipticals are more often found to be rotationally
supported.  Although the kinematics for ellipticals does not divide perfectly by
luminosity (Emsellem \etal 2011), the trend still exists.

The fact that some percentage of D ellipticals are part of the brightest ellipticals
suggests that a subset of D ellipticals are related to the cD class ellipticals
typical of brightest cluster members (BCM, Schombert 1986).  The cD class BCM's also
have shallower profiles and the highest luminosities, presumingly from a long history
of dynamical evolution where they have cannibalised lower mass companions, increasing
their luminosities and extending their envelopes due to higher velocity dispersions
from the energy of mergers (Duncan, Farouki \& Shapiro 1983; Schombert 1988; Oegerle
\& Hill 2001).  It is expected that an increase in the kinetic energy of the outer
stars will result in a shallow profile from simple kinematic arguments.  However,
only a 1/3 of the D ellipticals in our sample are in the highest luminosity category,
the remaining 2/3's cover a full range in terms of luminosity and are not in the same
luminosity bin as BCM's (i.e., the subset of ellipticals with the highest expected
merger rates).

Confirmation that at least some of the brightest D ellipticals are related to cD
class ellipticals comes from their local density.  A comparison of luminosity and
local density, $N$ (the number of galaxies within 1 Mpc), is found in Figure
\ref{mag_comp} along with an inset histogram of log $N$.  The poor correlation
between luminosity and local density is a well known mass segregation effect and
reflects the growth in luminosity of BCM's by dynamical evolution in a fashion that
does not directly reflect the local density (such as local velocity dispersion).
And, again as with luminosity, a subset of D ellipticals also tend to be found in
densest environments.  The D ellipticals in the densest regions also tend to be the
brightest, reinforcing their weak relationship with cD galaxies.  However, over 2/3's
of the D ellipticals are not in the highest density regions and are located in
regions similar to a majority of the normal ellipticals.  The deviations in profile
shape for these D ellipticals may be related to internal kinematics (altered by
external processes), but a mechanism will be required to produce D ellipticals but
leaving a majority of ellipticals unaffected.

\subsection{Structural and Isophote Properties}

\begin{figure}[!ht]
\centering
\includegraphics[scale=0.80,angle=0]{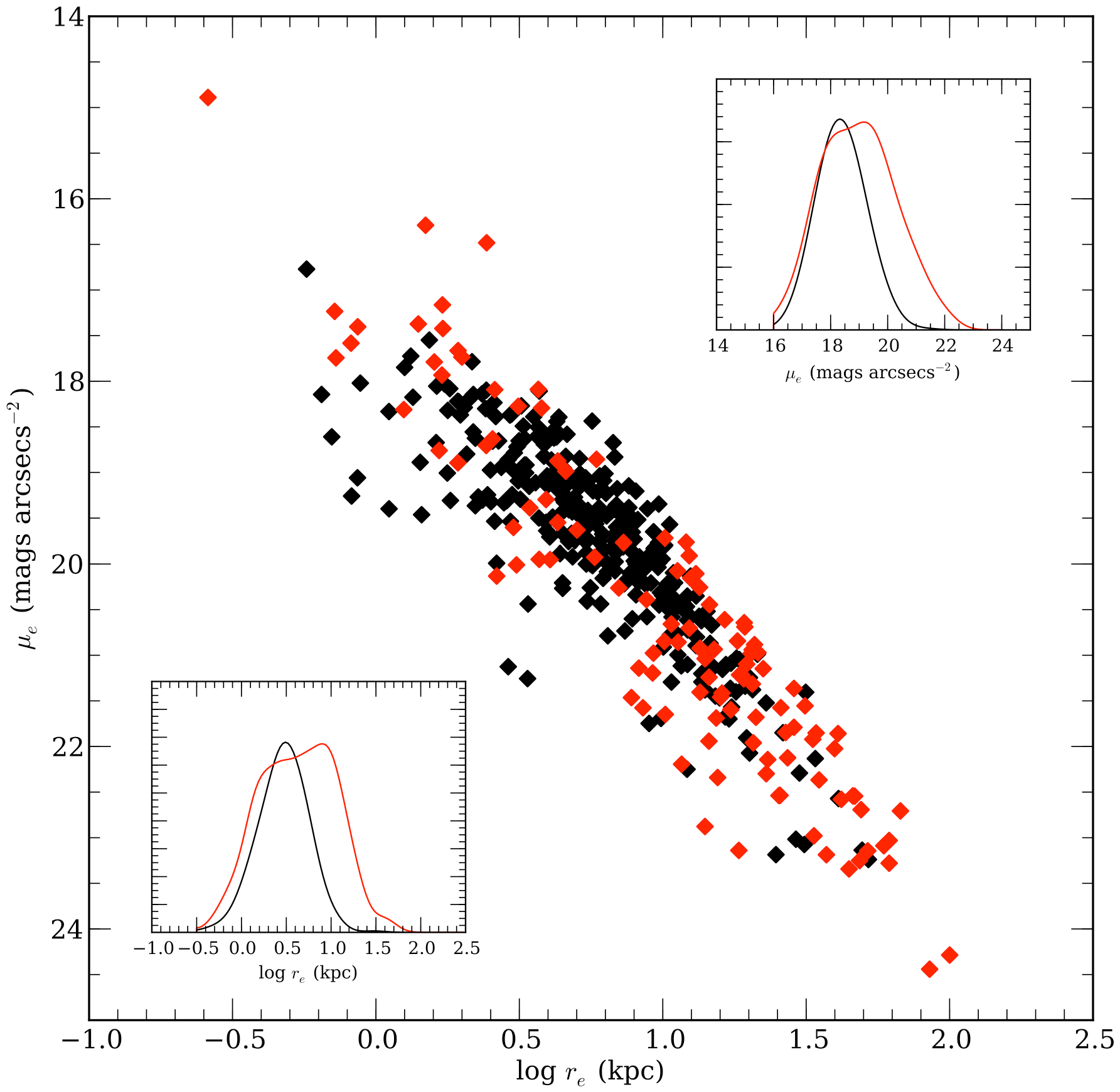}
\caption{\small Effective radius ($r_e$) versus effective surface brightness
($\mu_e$) from S\'{e}rsic $r^{1/n}$ fits.  Inset histograms display the scaled
normalized distributions of $r_e$ and $\mu_e$. Black symbols are normal ellipticals,
red symbols are D ellipticals.  The well known relationship between
scalelength and surface brightness is evident.  The D ellipticals cover the same range
in scalelength and surface brightness as ellipticals; however, they are more
concentrated at large radii reflecting their shallower slopes.
}
\label{el_comparison}
\end{figure}

Figure \ref{el_comparison} displays the structural comparison between normal
and D ellipticals using the effective radii and surface brightness from
S\'{e}rsic $r^{1/n}$ fits.  The D ellipticals, on average, have larger $r_e$ and
fainter $\mu_e$, as is expected from their shallower slopes.  However, as is a well
known problem with fitting functions, the different profile slopes do not reflect
into noticeable changes in the $r_e$ versus $\mu_e$ correlations.  The relationship in
Figure \ref{el_comparison} does not distinguish normal ellipticals from D ellipticals
other than the largest galaxies tend to be D ellipticals (in agreement with their
higher mean luminosities).

The D ellipticals display the extended scalelength and shallower profile slopes in a
similar manner as found for cD ellipticals.  However, cD ellipticals deviate
significantly from normal ellipticals in the $r_e$ versus $\mu_e$ diagrams (i.e.,
they are much shallower), presumingly a signature of past mergers that should be
common in the dynamical history of central cluster galaxies.  D ellipticals in our
sample appear to be simply an extension to the normal ellipticals $r_e$ versus
$\mu_e$ relationship.  While it is tempting to attribute some fraction of D
ellipticals as the result of strong dynamical growth in cluster cores, the remain
fraction have very similar scaling relations as normal ellipticals (given the
limitations of information extracted from fitting functions).

The D ellipticals, structurally, distinguish themselves primarily by profile slope.
The mean profile slope (measured between 17 and 23 $J$ mags arcsecs$^{-2}$) for
normal ellipticals is $-$2.1 (where luminosity density, $\Sigma$, goes as $\Sigma \propto
r^{-2.1}$).  The D ellipticals have a mean slope of $-$1.8.  Profile slope is a mild
function of luminosity (as can be seen in Figure \ref{templates}, the mean slope for
normals ellipticals ranges from $-$1.9 at $-$26 $J$ mags to $-$2.4 at $-$21.  Over
the same luminosity range, D ellipticals are consistently 0.2 more shallow in slope
at every luminosity bin.  This consistent difference is the reason that D ellipticals
are difficult to distinguish from normal ellipticals in Figure \ref{el_comparison},
for a shallow slope translates into a fainter surface brightness for a larger
effective radius, in nearly the same direction as the relationship for normal ellipticals.
Only a dramatic change in size, as seen for cluster BCM's, are detectable in the
$r_e$ versus $\mu_e$ diagram (see Figure 4 in Schombert 1987).

\begin{figure}[!ht]
\centering
\includegraphics[scale=0.80,angle=0]{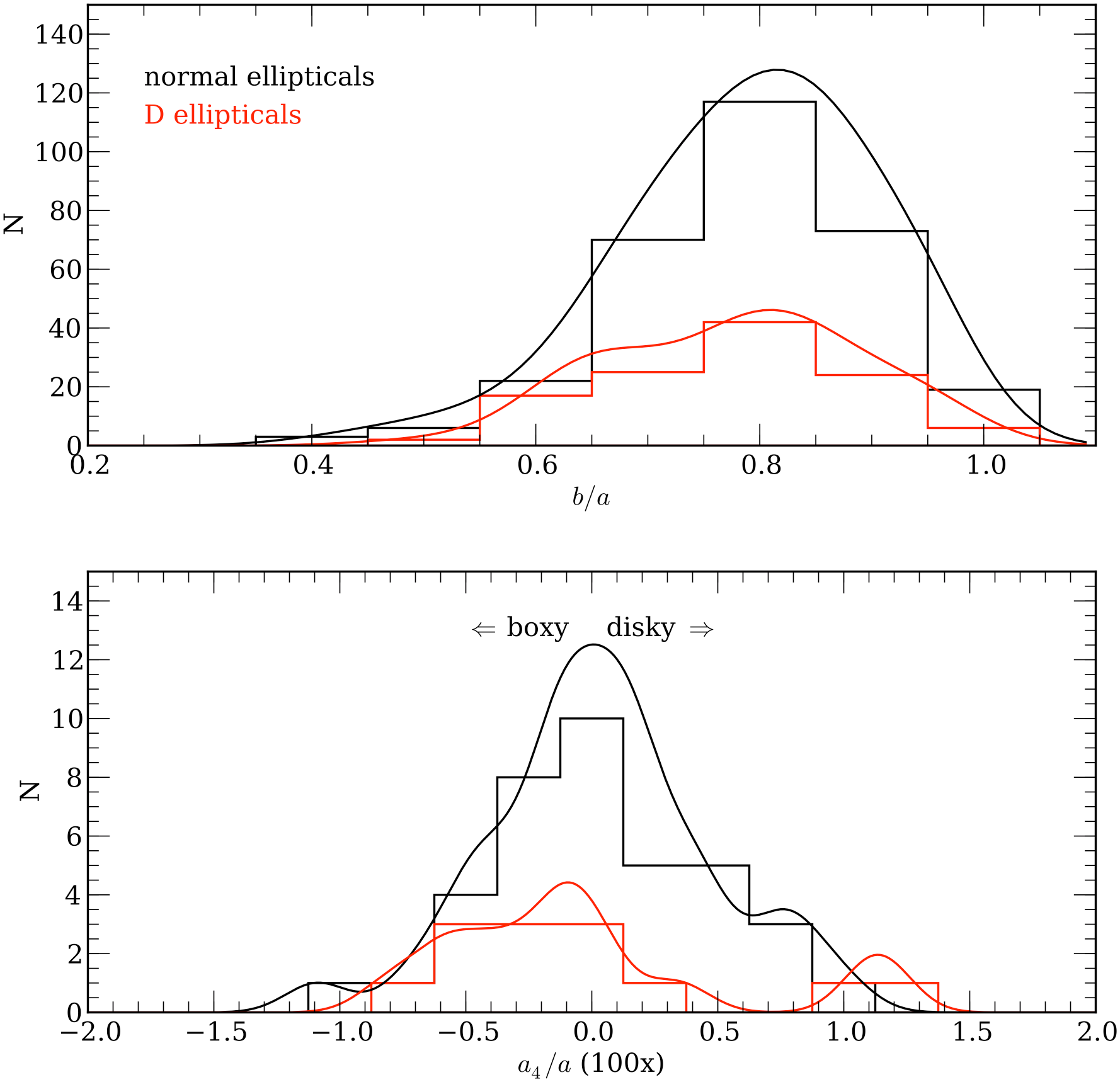}
\caption{\small Comparison of axial ratio ($b/a$) and isophote shape (as measured by
the fourth cosine coefficient, $a_4$) between normal and D ellipticals.  The
distribution of both isophotal measures is identical.  There is nothing in their
2D shapes to distinguish normal ellipticals from D ellipticals.
}
\label{isophotes_hist}
\end{figure}

Other correlations with structural characteristics were examined.  The top panel in
Figure \ref{isophotes_hist} displays the distribution of axial ratios ($b/a$) for
normal and D ellipticals.  The axial ratio is determined at the half-light radius
($r_h$), the empirical point where half the total luminosity is reached.  The
distributions are identical, there is no indication that D ellipticals are, on
average, flatter than normal ellipticals.  There is a slight increase in $b/a$ at 0.6
for D ellipticals suggesting that some fraction of D ellipticals may be misclassified
S0's, but this is less than 10\% of the sample.  Identical results are found for
$b/a$ value determined at 1/2$r_h$ and 2$r_h$.  Since S0's have mean $b/a$'s of 0.3
(Michard 1994), which is much flatter than the D ellipticals in our sample, it is
clear that the frequency diagram of axial ratios for D ellipticals is more similar to
normal ellipticals than S0's.

Kormendy \& Bender (1996) proposed that ellipticals be divided into two sequences
based on the shape of their isophotes (boxy versus disky).  Disky isophotes are
isophotes that are extended at the major axis and minor axis compared to an ellipse
of the same axial ratio.  They are, of course, common signatures in galaxies with
embedded oblate disks in a prolate or triaxial envelope (Scorza \& Bender 1995).
Boxy isophotes are flattened at the major and minor axis, taking on a box-like shape
compared to a best fit ellipse.  Disky galaxies dominate the low luminosity end of
the elliptical sequence which are often oblate in shape and whose kinematics are
dominated by rotation.  Boxy isophotes are a common feature in non-rotating
ellipticals with triaxial shapes dominated by anisotropic velocity distributions.
Boxy ellipticals are predominately higher in luminosity (Pasquali \etal 2007).
We investigated the occurrence of boxy and disky shaped isophotes for both types.
For example, disky isophotes may signal a flatter 3D shape for D ellipticals.

Our ellipticals were divided roughly into two types based on isophotal shape.
Unfortunately, the data reduction pipeline for 2MASS dithers the sky strip scans and
blurs the inner isophotes where boxy and disky shapes are usually detected (Schombert
2011).   In the end, we compared SDSS $g$ frames with the results from Bender \etal (1988),
the original study on isophote shapes.  We confirmed the fourth cosine coefficient
($a_4$) values for the Bender \etal sample from the isophotes of the SDSS $g$ frames
and found them all to be consistent with the original Bender \etal values.

The resulting $a_4/a$ values (taken from Bender \etal 1988) are found in the bottom
panel of Figure \ref{isophotes_hist}.  Again, the normal and D ellipticals have
identical $a_4/a$ distributions with the majority having isophotes are that purely
elliptical in shape.  Very few strongly boxy or strongly disk-like galaxies are found
in either type.  D ellipticals have a small number of galaxies with strongly disky
isophotes ($a_4/a > 1$); however, this small percentage has little statistical
significance.

In addition to isophotal shape, we also examined the change in the position angle of
the isophotal ellipse fits as a function of radius, known as isophotal twists
(Benacchio \& Galletta 1980).  Isophotal twists are used to probe the three
dimensional shape of ellipticals through statistical arguments.  The triaxial shape
for bright ellipticals, and oblate shape for faint ellipticals (deduced from
kinematic arguments, see below) are supported by isophotal twist analysis.  And,
while it is true that isophotal twists are more common in round galaxies (Galletta
1980; Nieto \etal 1992), and rare in flattened systems, part of this effect is due to
the difficulty in assigning a position angle to a very round isophote.

Much like the results for isophotal shape, the distribution isophotal twists was the
same for the elliptical and D elliptical samples.  There was no indication that D
ellipticals had fewer position angle changes, signaling an oblate shape, versus
normal ellipticals.  The nature of the different profile shape for D ellipticals
compared to ellipticals is not revealed by any characteristic related to the 3D mass
density shape.  In addition, visual examination of the profile subtracted images
found no evidence for peculiar features, such as tidal tails or shells, compared to
the normal ellipticals sample.

\subsection{Kinematics}

Of most interest is whether there is a kinematic signature to distinguish the normal
ellipticals from D ellipticals.  The expectation for a difference is low since D
ellipticals cover the same range in galaxy mass (i.e., luminosity) as the normal
ellipticals and, therefore, are presumed to follow the same trends in kinematics.
Also, the extended profile shapes of the D ellipticals occurs at radii much larger
than sampled by kinematic studies.  Thus, there is no reason to believe that internal
kinematics will be correlated with kinematics in the envelopes responsible for outer
structure.

\begin{figure}[!ht]
\centering
\includegraphics[scale=0.80,angle=0]{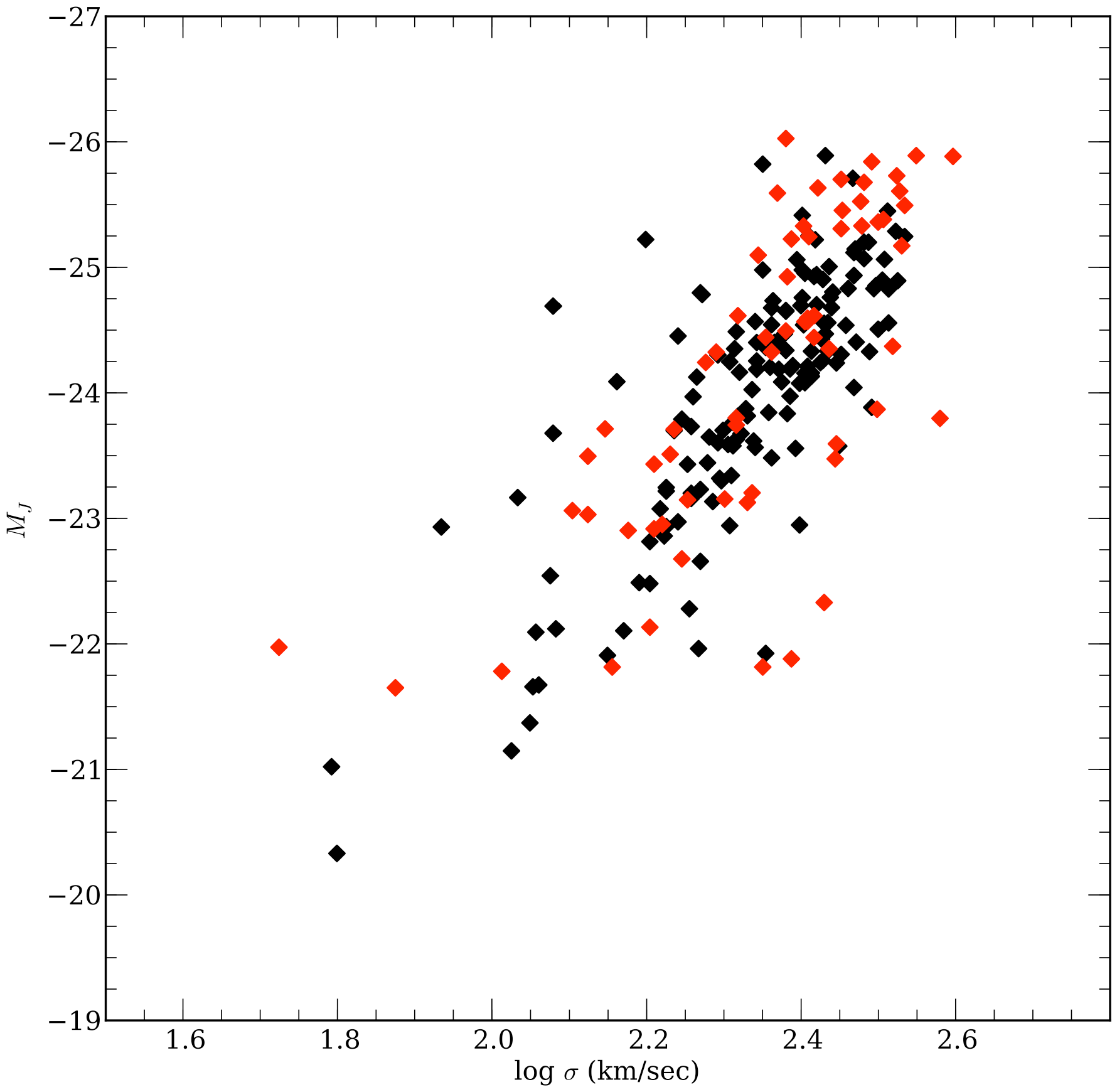}
\caption{\small A comparison between normal and D ellipticals in the Faber-Jackson
diagram, a plot of velocity dispersion versus total luminosity.  Black symbols are
normal ellipticals, red symbols are D ellipticals.  Other than a slight concentration
at high velocity dispersion, the relationship between normal and D ellipticals is the
same, although the central velocity dispersion does not measure the kinematics of the
outer envelope where the structural difference between normal and D ellipticals
occurs.
}
\label{vel_mag}
\end{figure}

Unfortunately, although the photometric sample is large, the kinematic information
for these same early-type galaxies is limited.  As a first comparison, we have
plotted in Figure \ref{vel_mag} the velocity dispersions from the SDSS database
(Bernardi \etal 2003) versus their total $J$ luminosities.  There were velocity
measurements for 50\% of the sample, evenly divided between normal and D ellipticals.
As can be seen in Figure \ref{vel_mag}, there is little difference in the
relationship between velocity dispersion and luminosity (i.e., stellar mass) for
normal and D ellipticals.  The normal ellipticals are slightly more correlated than
the D ellipticals.  The D ellipticals have a slightly higher dispersion and are more
concentrated at higher luminosities (but slightly lower velocity dispersion).
Although this is not surprising as the velocity dispersion versus luminosity relation
even for S0's is identical to ellipticals and lacks a discriminator capable with
respect to morphology (Dressler \& Sandage 1983).

\begin{figure}[!ht]
\centering
\includegraphics[scale=0.80,angle=0]{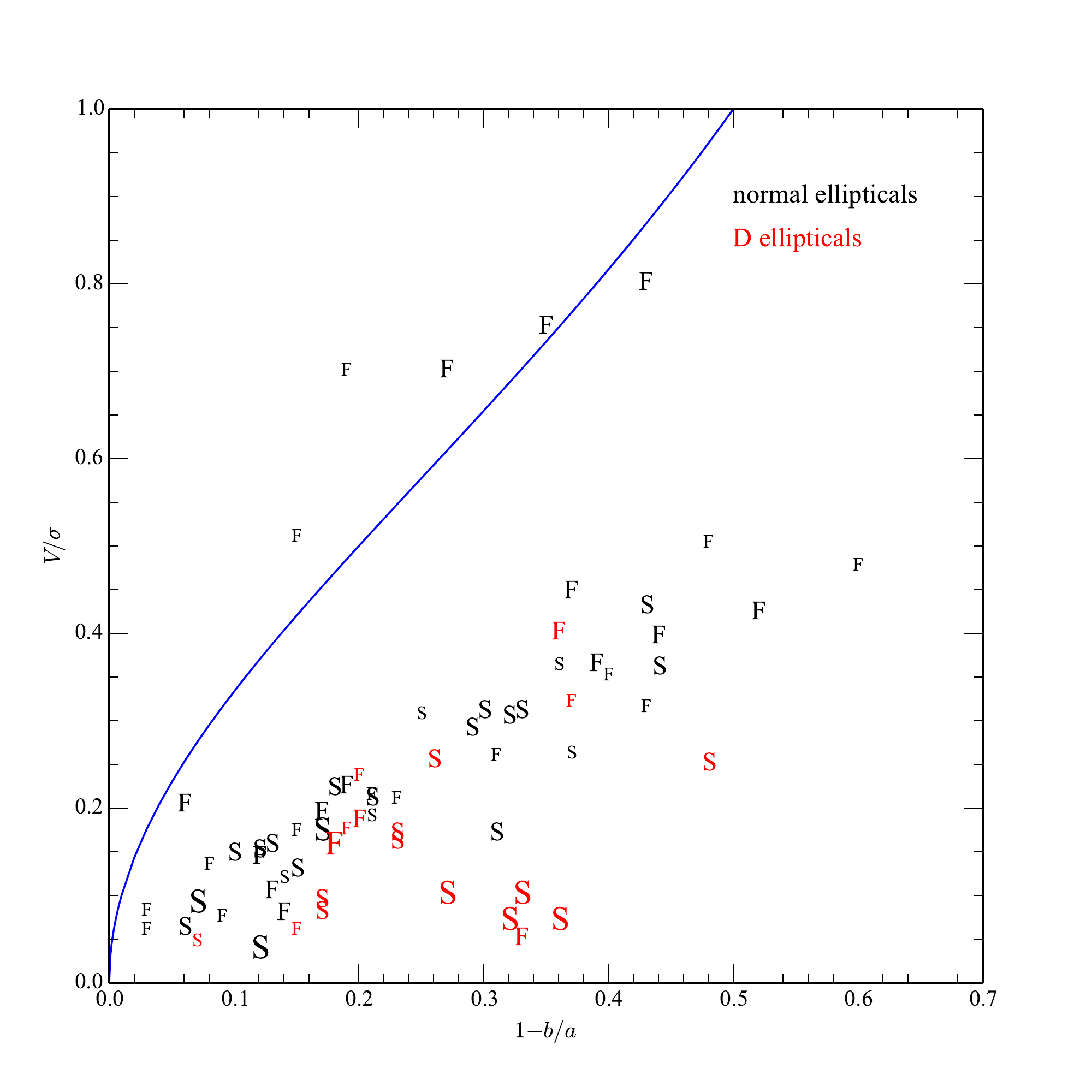}
\caption{\small The anisotropy diagram for normal (black) and D (red) ellipticals.
The anisotropy diagram is a plot of the ratio of rotational velocity maximum ($V$) to
the central velocity dispersion ($\sigma$) versus the galaxy's ellipticity ($1-b/a$).
Oblate galaxies follow the blue curve.  Galaxies with prolate or triaxial shapes fall
below the curve.  Each datapoint is labeled by their SAURON designation of fast or
slow rotators.  The size of the symbol is proportional to the total luminosity of the
galaxy.  The D ellipticals avoid the oblate curve (although very few normal
ellipticals are near this shape as well).  And D ellipticals tend to have the lowest
$V/\sigma$ values, with a grouping of flattened D ellipticals below the normal
elliptical trend line.
}
\label{V_sigma}
\end{figure}

Any important kinematic signature would probably be related to rotation, not velocity
dispersion.  For dissipation leads to stronger rotation and strong mergers decrease
the importance of rotation (Barnes 1989).  The mean diagnostic for the underlying
kinematics in ellipticals is the $V/\sigma$ parameter, the ratio between the maximum
rotation speed and the velocity dispersion (Binney 2005), usually plotted against
galaxy ellipticity (the so-called the anisotropy diagram).  Searching the literature,
we have taken data for 68 galaxies in our sample (49 normal ellipticals, 19 D
ellipticals) from Davies \etal (1983) and Emsellem \etal (2011).  The resulting
anisotropy diagram is shown in Figure \ref{V_sigma}.

The anisotropy diagram indicates the underlying 3D shape of an elliptical where the
blue line in Figure \ref{V_sigma} is the canonical relationship for an oblate
galaxy.  Data points below this curve represent prolate and triaxial galaxies.  In
Figure \ref{V_sigma} the size of the symbol is proportional to the luminosity of
the galaxy and each symbol is marked as either 'F' or 'S' as a designation of the
SAURON project's classification as a fast or slow rotator (Emsellem \etal 2007).  As
can be seen in Figure \ref{V_sigma}, a majority of the normal and D ellipticals
lie well below the oblate curve, despite indications of fast or slow rotation.  The
few normal ellipticals near the oblate line are all fast rotators, indicative of a
rotational supported oblate shape.  The D ellipticals are mostly slow rotators with
low $V/\sigma$ values, although not dramatically different than the distribution of
normal ellipticals.  We note that the brightest D ellipticals have the lowest
$V/\sigma$ values for their apparent ellipticity.

\section{Conclusions}

It is somewhat surprising that elliptical structure is as smooth a function of
luminosity as displayed by the isophotal radius relations.  For even normal
ellipticals display a range of underlying kinematics that reflect many components
(Emsellem \etal 2011).  If kinematics dominate the structure of a galaxy, as
reflected in its surface brightness profile, then the structure of ellipticals should
take on a wide variety of shapes and slopes (although we note that most kinematic
studies are confined to the core regions and kinematic statements about the outer
regions are uncertain).  We can only be guided by the fact that numerical simulations
that invoke the two most common formation scenarios (dissipational monolithic
collapse, Nipoti \etal 2006 and dissipationless merging, Aceves \etal 2006, Naab \&
Trujillo 2006) both result in S\'{e}rsic and $r^{1/4}$ shapes.  In other words,
simulations indicate that complicated kinematics still result in smooth surface
brightness profiles due to a variety of relaxation processes that produces present-day
elliptical galaxies in a state of quasi-equilibrium.  

It is also interesting to note that the transition from rotation dominated kinematics
and an anisotropic or pressure support kinematics occurs at approximately $M_J=-$23,
which is also the structural point where normal elliptical structure transitions from
S\'{e}rsic shapes to nearly $r^{1/4}$ in shape.  As can been seen in the templates in
Figure \ref{templates}, faint elliptical structure is not $r^{1/4}$ in shape in that
it has too much downward (fainter) curvature at small and large radii (which can be
captured by the S\'{e}rsic $n$ index).  For a galaxy brighter than $M_J=-23$, the
profile shape deviates from a strict $r^{1/4}$ shape, but in a fashion predicted by
the violent relaxation models of Hjorth \& Madsen (1991).  Comparison to their Figure
3 for the residuals from a $r^{1/4}$ shape is remarkably similar to the deviations
from our templates (see our Figure \ref {fit_residuals}), although the their
simulations are not unique as a wide range of initial conditions result in similar
galaxy shapes.  However, it does suggest that a history of dry mergers (without
dissipation) plays some role in the formation of bright ellipticals.

With the construction of the templates, and sequential re-comparison to all the
elliptical profiles, comes the discovery that 1/3rd of the ellipticals in our sample,
all classified as pure ellipticals based on visual morphology, deviate in a
systematic fashion from the normal ellipticals templates.  These ellipticals have
shallower profiles than their templates at their respective luminosities, which would
give them a diffuse appearance on a photographic plate.  Thus, we refer to these
objects as D ellipticals in acknowledgement of the pre-existing diffuse designation
from Matthew, Morgan \& Schmidt (1964), although this usually applied to 1st ranked
galaxies in rich clusters (Schombert 1988).  This dichotomy into normal (template)
and D elliptical is unrelated to the core versus coreless separation, as this
distinction is confined to the inner 1-2 kpc, nor is it related to the boxy versus
disky isophotal shape for ellipticals (see \S 5.3).

The D ellipticals do not distinguish themselves, radically, from normal ellipticals
through any physical characteristic other than structure.  While they tend to be
brighter and located in denser regions of the Universe, they cover the same
luminosity and local density space as normal ellipticals.  Their structural
properties, as derived from fitting functions, have the same relationships as normal
ellipticals (although this is more a statement concerning the limitations of fitting
functions as D ellipticals clearly deviate in their profile shape).  The isophotal
characteristics of normal and D ellipticals are identical in terms of their axial
ratio and 2D isophote shape.  Their kinematics are similar, although none of the D
ellipticals display strong rotation signatures.

Another key point is that a template pattern could not be constructed for D
ellipticals, although their numbers are 1/3rd that of the normal ellipticals in the
sample and the template algorithm may not have converged due to small numbers.  In
other words, even though the profiles of D ellipticals are shallower than normal
ellipticals, the profile slope is not a smooth function of luminosity.  If
small numbers are not to blame, then the lack of uniformity to the D elliptical
profiles suggests that these galaxies are formed by random or stochastic processes.

From the connection between the D ellipticals and the cD class found in rich
clusters, we speculate that D ellipticals are more diffuse than normal ellipticals
due to a recent history of dry (i.e. dissipationless) mergers.  The D ellipticals are
slightly more common a high luminosities and rich galaxy environments.  Recent
mergers would be more common in high density regions, and result in brighter
ellipticals.  While a tenuous connection, mergers have the right signature (more
energy added to the stars in the outer orbits producing a more extended profile) and
the degree of randomness induced by mergers would explain the lack of correlation
with any other physical property to the D ellipticals.

As many galaxy formation scenarios have all ellipticals forming from mergers
(Kauffmann \& Charlot 1998), then the next concern is whether D ellipticals are stable
or a transition class of objects.  If stable, and all ellipticals are the result of
mergers, than some special circumstances exist for D ellipticals with respect to
normal ellipticals.  If D ellipticals evolve into normal ellipticals, then the orbits
of the stars after a recent merger which produces a D elliptical profile must later
stabilize into a normal elliptical shape as given by the templates in Figure
\ref{template}.  If mergers are common for all ellipticals, than does the roughly
1-to-3 ratio of normal ellipticals to D ellipticals imply a stabilization timescale,
or a current merger rate.

As evidence of dissipation formation is found in many ellipticals (e.g., color
gradients), perhaps the difference between normal and D ellipticals is a measure of
the importance of wet mergers, with dissipation effects, to later dry mergers where
violent relaxation effects dominate.  As strong dissipation leads to isotropic
velocity distribution (Navarro 1990), and hierarchical mergers lead to structural
non-homology (Dantas \etal 2003), the near homology of normal ellipticals implies a
strong component of wet mergers in the early epochs where collapse prefers homology.
Then, later dry mergers indice mild structural non-homology.

With respect to the origin of D ellipticals, it is already known that high redshift
ellipticals lack the outer envelopes (Szomoru \etal 2012), which implies that most
elliptical galaxies start forming their stars at high redshift in a dissipative
environment, rapidly become very massive by $z = 2$ by later mergers (Keres et al.
2009; Feldmann et al. 2011; Oser et al. 2012).  But that the formation of the outer
envelope occurs in an era dominated by dissipationless mergers, where new baryonic
matter is added to the outer parts of the galaxies over time with very little star
formation, (van Dokkum et al. 2010; Saracco et al. 2012).  This form of rapid
structural evolution promotes the growth of the outer envelope with very little
change to the central regions, identical to the differences we detect between normal
and D ellipticals.

Under hierarchical scenarios, mergers play a major roles in galaxy formation.  They
are expected to happen frequently and provide a natural way to increase the size of a
galaxy.  The addition of stellar material, particularly at large radii, cause the
luminosity distribution to change significantly resulting in a significant increase
of the S\'{e}rsic index (Hilz \etal 2012).  When the relaxation period ends, stellar
energy is not exchanged and positive energy stars escape and loosely bound stars
expand to larger radii.  This provides a natural mechanism to explain the changes in
the shape of the templates as a function of total luminosity, and their similarity to
shapes predicted by violent relaxation (Hjorth \& Madsen 1995).

However, different types of dry mergers predict different profile shapes and
kinematics.  For example, equal mass mergers lead to anisotropic kinematics and
shallower profiles (that increases with the galaxy masses) while unequal mass mergers
(e.g., accretions) imparts rotation and more concentrated profiles (Khochfar \&
Burkert 2005).  Numerical cosmological simulations find that the mass assembly of
ellipticals is dominated by accretion of small galaxies with mass-ratios near 1-to-5
(Oser \etal 2012; Lackner \etal 2012).   If D ellipticals are the
result of nearly equal mass mergers, than their lower numbers compared to normal
ellipticals is in agreement with the expectations from these simulations.  Therefore,
we propose that normal ellipticals are the result of late dry mergers will small
companions, while the shallower D ellipticals are the result of recent dry mergers
with nearly equal mass companions.

This may provide a natural mechanism for the division of rotation kinematics in
ellipticals into fast and slow rotators.  The SAURON project (Emsellem \etal 2007)
finds all fast rotators to be low luminosity, but slow rotators, although brighter in
the mean, are found at all luminosities, like D ellipticals.  Slow rotators may be
result of dissipationless mergers, where most of the baryonic momentum is expelled
outward resulting in diffuse envelopes.  Thus, the expectation that all D ellipticals
be triaxal, as seen in Figure \ref{V_sigma}.

There are several testable prediction from the above scenario for the formation of D
ellipticals.  For example, mergers can produce gradients and color-magnitude relation
(Kauffmann \& Charlot 1998); however, there should be measurable differences between
the gradients in normal and D ellipticals.  Structural non-homology can be driven by
varying star formation histories (Bekki 1998), so age and metallicity gradients would
test the levels of star formation during the past mergers.  Clearly, the kinematics
of D ellipticals envelopes should be more energetic than normal ellipticals, but this
would require deep optical spectroscopy of their envelopes, perhaps a future project
for our next generation ground-based telescopes (Raskutti, Greene \& Murphy 2014).

\noindent Acknowledgements: 

The software for this project was supported by NASA's AISR and ADP programs.  This
publication makes use of data products from the Two Micron All Sky Survey, which is a
joint project of the University of Massachusetts and the Infrared Processing and
Analysis Center/California Institute of Technology, funded by the National
Aeronautics and Space Administration and the National Science Foundation. In
addition, this research has made use of the NASA/IPAC Extragalactic Database (NED)
which is operated by the Jet Propulsion Laboratory, California Institute of
Technology, under contract with the National Aeronautics and Space Administration. 

\pagebreak

\begin{deluxetable}{lccccc}
\tablecolumns{6}
\small
\tablewidth{0pt}
\tablecaption{$r^{1/4}$ and S\'{e}rsic $r^{1/n}$ template fits}

\tablehead{
\colhead{} & \multicolumn{2}{c}{$r^{1/4}$} & \multicolumn{3}{c}{S\'{e}rsic $r^{1/n}$}  \\
\cline{2-3} \cline{4-6} \\
\colhead{$M_J$} &
\colhead{log $r_e$} &
\colhead{$\mu_e$} &
\colhead{log $r_e$} &
\colhead{$\mu_e$} &
\colhead{log $n$} \\

\colhead{} &
\colhead{(kpc)} &
\colhead{} &
\colhead{(kpc)} &
\colhead{} &
\colhead{} \\

}

\startdata

\cutinhead{Inner Fits}

$-$21.5 & 0.23 & 19.05 & 0.22 & 18.97 & 0.22 \\
$-$22.0 & 0.33 & 19.09 & 0.30 & 18.94 & 0.41 \\
$-$23.0 & 0.52 & 19.18 & 0.52 & 19.17 & 0.62 \\
$-$24.0 & 0.65 & 18.97 & 0.65 & 19.02 & 0.64 \\
$-$25.0 & 0.82 & 19.07 & 0.77 & 18.84 & 0.54 \\

\cutinhead{Outer Fits}

$-$21.5 & 0.42 & 14.96 & 1.89 & 19.28 & 0.08 \\
$-$22.0 & 0.99 & 17.14 & 2.25 & 19.22 & 0.25 \\
$-$23.0 & 2.99 & 18.93 & 4.04 & 19.60 & 0.38 \\
$-$24.0 & 5.09 & 19.24 & 5.87 & 19.51 & 0.44 \\
$-$25.0 & 8.70 & 19.62 & 8.65 & 19.58 & 0.54 \\

\cutinhead{Full Fits}

$-$21.5 & 0.85 & 17.45 & 1.72 & 19.02 & 0.13 \\
$-$22.0 & 1.38 & 18.16 & 2.02 & 18.95 & 0.30 \\
$-$23.0 & 3.18 & 19.10 & 3.25 & 19.09 & 0.49 \\
$-$24.0 & 4.72 & 19.07 & 4.79 & 19.06 & 0.53 \\
$-$25.0 & 7.58 & 19.29 & 7.49 & 19.25 & 0.58 \\

\enddata
\end{deluxetable}


\end{document}